\documentclass[natbib]{svjour3}             
\smartqed  
\usepackage{graphicx}
 \usepackage{mathptm}      
 \usepackage{aps-bibstyle}  
 \usepackage{times}
 \usepackage{textcomp}
 \usepackage{color}
\definecolor{red}{RGB}{255,0,0}

\begin{document}

\title{Solar magnetoconvection and small-scale dynamo}

\subtitle{Recent developments in observation and simulation}


\author{J. M. Borrero$^1$ \and 
        S. Jafarzadeh$^2$ \and
	M. Sch{\"u}ssler$^3$\and
	S. K. Solanki$^{3,4}$
	}

\authorrunning{J. M. Borrero et al.}

\institute{$^1$ Kiepenheuer-Institut f{\"u}r Sonnenphysik,
                Sch{\"o}neckstr 6, 79104 Freiburg, Germany\\
           $^2$ Institute of Theoretical Astrophysics, University of
                Oslo, P.O. Box 1029 Blindern, N-0315 Oslo, Norway\\
	   $^3$ Max-Planck-Institut f{\"u}r Sonnensystemforschung, 
                Justus-von-Liebig-Weg 3, 37077 G{\"ottingen}, Germany\\
	   $^4$ School of Space Research, Kyung Hee University, 
                Yongin, Gyeonggi-Do,446-701, Republic of Korea
}

\date{\today}

\maketitle

\begin{abstract}
A number of observational and theoretical aspects of solar
magnetoconvection are considered in this review. We discuss recent
developments in our understanding of the small-scale structure of the
magnetic field on the solar surface and its interaction with convective
flows, which is at the centre of current research. Topics range from
plage areas in active regions over the magnetic network shaped by
supergranulation to the ubiquituous `turbulent' internetwork fields.
On the theoretical side, we focus upon magnetic field generation by
small-scale dynamo action. 
\end{abstract}

\section{Introduction}
\label{intro}
New observational facilities and instruments (on the ground and in
space), innovative methods for data analysis, and the rapid growth of
computing power have led to considerable progress in our understanding
of solar magnetoconvective processes during the last decade. This ranges
all the way from the smallest observable magnetic structures in `quiet'
areas to the dynamical fine structure of sunspot umbrae and penumbrae.
The interplay of observations and radiative 3D-MHD simulations is
crucial for this progress: simulations permit us to identify the
physical processes leading to the observed phenomena once their
consistency with the observations is shown by comparing `synthetic' with
real observations. The simulations then can sharpen the questions
addressed by the observations by predicting further properties of the
magnetic, thermodynamic, and flow structure.

In this review, we consider a subset of the solar magnetoconvective
phenomena, also focussing on more recent developments. Plage and network
fields are of considerable interest, not only because of their intrinsic
importance for understanding the interaction of magnetic field and
convective flows at various spatial scales, but also because of their
crucial role for solar irradiance variations with potential impact on
terrestrial climate variations. Internetwork fields in the `quiet' Sun
probably represent a huge reservoir of restless `turbulent' magnetic
flux, possibly heralding a fully magnetized state (in the sense of
equipartition between magnetic and kinetic energy of the convective
motions) of the whole convection zone with important implications for
the solar dynamo and the generation of differential rotation. A
plausible mechanism for the generation of the internetwork field is
small-scale dynamo action by a flow with chaotic streamlines, a truly
`turbulent' dynamo. Numerical simulations suggest that such a dynamo
could be active throughout the convection zone, but these simulations
can only be run for effective Reynolds numbers and magnetic Prandtl
numbers far away from solar values. Nevertheless, state-of-the-art
radiative 3D-MHD simulations provide results which are consistent with
observations.  In what follows, we give an overview of the state of
research in the three areas mentioned above: plage and network fields
(Sect.~\ref{sec:plage}), internetwork fields
(Sect.~\ref{sec:internetwork}), and small-scale dynamo
(Sect.~\ref{sec:ssd}).

\section{Plage and network fields}
\label{sec:plage}

\subsection{General properties of plage and network fields}

At photospheric layers active regions are mainly composed of large, cool
magnetic features such as sunspots and pores and smaller, bright magnetic
elements forming the plage regions, named after the bright appearance of such
regions at chromospheric heights. They are sometimes also called facular
fields after the bright faculae visible mainly near the limb in white light
(or in other photospheric continuum radiation). Similarly, network features are
named after the network of brightenings seen in the cores of chromospheric
lines and covering the quiet Sun.

Plage and network magnetic fields are intermediate between the large and
dark sunspots and pores on the one hand, and the small weak internetwork
magnetic features on the other hand. They are distinguished from the
pores by the fact that they are considerably brighter in white light,
being on average roughly the same brightness as the average quiet Sun
(around solar disc centre), or brighter (near the limb). Their larger
magnetic fluxes and their arrangement inside active regions or at
the boundaries of supergranule cells distinguishes them from the
internetwork elements.

Earlier reviews covering the physics of photospheric flux tubes and in
particular plage and network magnetic fields have been written by
\citet{Solanki1993,Stenflo1994,Solanki2006,deWijn2009}; see also
\citet{Stein2012} for a review of magnetoconvection, which is central to the
physics of faculae and the network, and
\citet{Wiegelmann2014} for a more recent overview of magnetic fields in the
solar atmosphere, but focusing more on the corona.

Plage and network magnetic fields are distinguished from each other by the
fact that the density of magnetic flux typically present inside plage is higher, and by
their location, with the former being found inside active regions, while network fields are
present all over the Sun (including among the decay products of
active regions, which form the so-called enhanced network), but concentrated
mainly at the edges of supergranules with a length scale of 15--30~Mm. This implies that
the plage fields are restricted to the activity belts, i.e. to latitudes lower
than roughly $\pm 30^\circ$ (see \citealt{Hale1925}), while the network is
seen at all latitudes \citep{Muller1994b}. In addition, plage and network show
a markedly different
behaviour over the solar activity cycle, with the solar surface area covered by
plage regions waxing and waning strongly in phase with the number of sunspots, while
the network fields display a much weaker variation with time. Indeed, there
is some controversy about the extent to which network fields vary in phase or
in antiphase with the solar cycle (\citealt{Harvey1993,Harvey1993b,Hagenaar2003}). \citet{Jin2011} have proposed that the phase vs. anti-phase behaviour
depends on the size and flux of the network magnetic features, with the
larger features varying in phase, while the smaller features vary in
antiphase with the sunspot cycle.

Also, plage and network fields have partly different origins. A large fraction of the plage fields emerge in place, but some are the decay products of the sunspots in
their active regions \citep{Petrovay1999}. Therefore, plage fields are generally produced by the global dynamo. The network in turn is partly the result of
the decay of active region plage. To a larger extent, however, the network is fed by
ephemeral regions, also called ephemeral active regions, small bipolar regions without
sunspots with magnetic fluxes of roughly $10^{18}$--$10^{20}$~Mx \citep{Dodson1953,Harvey1973a,Harvey1973b}.

The constant emergence of new magnetic flux in the form of ephemeral regions
over the solar surface, at a rate of $3\times10^{21}$ Mx per hour (assuming a
homogeneous emergence rate over the entire Sun; \citealt{Schrijver1997}),
implies a constant
removal of magnetic flux at the same rate. This has led to estimates of the
flux turnover time in the magnetic network of 40~h by \citet{Schrijver1998}
and 14~h by \citet{Hagenaar2001}.

Recently, the importance of the contribution to the network by magnetic
flux emerging in the intranetwork has been pointed out
\citep{Gosic2014}. Although both, merging and cancellation processes
take place \citep{Iida2012}, the former was found to strongly dominate,
resulting in a rate of net flux transfer of $1.5\times10^{24}$~Mx per
day from the internetwork to the network.  \citet{Gosic2014} argue that
this implies internetwork fields replace the entire flux in the network
within 18-24 hours, on the same order as the replacement time deduced by
\citet{Hagenaar2001} based on ephemeral regions. The authors do not
discuss the pressing question of why, if the internetwork flux in the
two polarities is well balanced while the network flux in their
observations clearly is not, merging dominates so strongly over
cancellation. Also unclear is what happens with the internetwork flux of
the opposite polarity, which is not merged with the network. Also
unclear is why, if the flux contributed by the internetwork to the
network is an order of magnitude larger than that contributed by
ephemeral regions, the turnover times for the network flux found by
\citet{Hagenaar2001} (based only on ephemeral regions) is similar to
that obtained by \citet{Gosic2014} (including also internetwork
features). This could imply that the network in the region considered by
\citet{Gosic2014} carried a particularly large amount of flux.  
Furthermore, the decay or fragmentation of network elements provides an
additional source of internetwork flux, so that the net contribution of
the internetwork to the network field could actually vanish.

\subsection{Observations of magnetic concentrations making up plage and the network}
\label{subsect:obs_plage_network}

In photospheric layers, where most of the measurements of solar magnetic
fields have been made, both, plage regions and the network are composed of
groups of more or less discrete magnetic flux concentrations. These
concentrations display a range of sizes \citep[e.g.][]{Utz2009,Feng2013} and
magnetic fluxes, whose probability distribution function follows a
power law with an exponent of $-1.85$ according to \citet{Parnell2009}, or
$-2$ according to \citet{Harvey1993b}. These values indicate that large and
small magnetic
concentrations contribute roughly the same amount to the total unsigned magnetic
flux present at any given instant on the solar surface (with the larger features
dominating if the value of \citealt{Parnell2009} is used). Owing to
the much shorter lifetime of ephemeral regions, this approximate equality
implies they bring magnetic flux to the solar surface at a much higher rate
than the larger active regions. \citet{Zirin1987} has estimated that there is a
factor of 100 difference in the fluxes emerging over a solar cycle in these
two types of bipolar features (cf. \citealt{Thornton2011,Zhou2013}).

Magnetic fields in active regions and the network are mainly strong, i.e. on the order of
1-2 kG \citep{Stenflo1973,Wiehr1978,Dara-Papamargaritis1983,Rabin1992a,Rabin1992b,Rueedi1992,Grossmann-Doerth1996,Martinez-Pillet1997,Sanchez-Almeida2000,Viticchie2011,Buehler2015},
also see \citet{Blanco-Rodriguez2010} and \citet{Kaithakkal2013} for polar
faculae. These fields are close to vertical
\citep[e.g.,][]{Sanchez-Almeida1994,Bernasconi1995,Martinez-Pillet1997,Buehler2015};
cf. \citet{Leighton1959}, albeit with limitations of their instrument at the time.
Note that the field strength is a strong function of height,
dropping from high values of around 1500-2000 G in the deepest observable layers
\citep{Rabin1992a,Rueedi1992,Buehler2015} to a field
strength of 250-500 G in the upper photosphere \citep{Zirin1989,Bruls1995,Buehler2015};
see also \citet{Moran2000}. In addition to the kG flux concentrations, small patches of transient horizontal
field are also found in active regions \citep{Ishikawa2008,Ishikawa2009}, with
properties that are similar to corresponding patches in the quiet Sun. In
particular, the smaller ones are isotropically oriented, although the larger
ones appear to have a slight preference for the orientation of the active
region.

According to \citet{Abramenko2005} magnetic flux concentrations within
active regions possess fluxes of $10^{18}$--$10^{20}$~Mx, whereby features
with fluxes larger than $10^{19}$~Mx are likely darker than the surroundings
\citep[e.g.,][]{Grossmann-Doerth1994}. The network is composed of
smaller magnetic features hosting less magnetic flux, with kG flux tubes
carrying a flux of around $10^{17}$~Mx being the smallest such entities
that have been isolated so far. Even smaller magnetic features
are regularly seen in radiation MHD simulations and are also likely to be present
on the Sun. \citet{Solanki1999} pointed out that magnetic features with magnetic fluxes differing by
up to 6 orders of magnitude have roughly the same field strength of
1000-2000~G when averaged over their cross-sections. Such magnetic features
range from the
smallest kG flux tubes (magnetic elements) to large sunspots. Note that only the
field strength averaged over the whole cross-section remains roughly
independent of magnetic flux, the
field strength at the core of the features changes by a considerable amount.
Magnetic features with flux below a few times $10^{16}$~Mx have weaker fields
\citep[e.g.][]{Solanki1996,Khomenko2003,Orozco2008,Ishikawa2009,Stenflo2011,Utz2013}.
Note that the field strength of individual magnetic structures may also be a
function of time, e.g. emerging magnetic flux is usually associated with a weaker field
\citep[e.g.][]{Zwaan1985,Zhang1992,Lites1998,Centeno2007,Martinez-Gonzalez2009,Cheung2014},
but in small features the field strength may also vary significantly after that \citep[e.g.][]{Martinez-Gonzalez2011,Requerey2014}.

The strong magnetic fields in plage decrease the convective blueshift of
spectral lines \citep{Livingston1982,Cavallini1985,Brandt1990}
in agreement with the smaller upflow velocities in granules along with the
faster downflows in intergranular lanes seen by \citet{Kostik2012}.
These authors also find that the convective flows reach greater heights in
facular areas.
The contrast of the granulation and the sizes of individual granules become
smaller in plage while their lifetimes increase, resulting in what is called abnormal
granulation~\citep{Title1992,Berger1998,Narayan2010}.

\citet{Martinez-Pillet1997} noted from Stokes
vector observations with a spatial resolution of around 1~arc sec obtained
with the Advanced Stokes Polarimeter (ASP) that the magnetic concentrations
in plage displayed no significant downflows beyond about 250~m/s, confirming
a result first found by \citet{Solanki1986}. The absence of
downflows was later challenged by \citet{Bellot-Rubio2000}, who argued
that the tiny wavelength shifts of the many lines in a Fourier-Transform-Spectrometer
(FTS) Stokes $V$ spectrum could only be interpreted in terms of a strong downflow
inside the magnetic features. This was in turn criticised by
\citet{Frutiger2001}, who showed that this signature could also be produced
by velocities not requiring a net flow. Consequently, spatially
unresolved data have not been able to provide a definitive answer to the
question of flows within magnetic elements. Upflows of
around 1-2 km/s inside plage magnetic elements, surrounded by 1.5-3.3~km/s
downflows, were reported by \citet{Langangen2007}. Recently, \citet{Buehler2015}
have confirmed the location of the magnetic concentrations inside the downflow lanes
separating granules, but without harbouring strong downflows themselves. At
individual locations up- and downflows within the magnetic features are
present, but in an average sense they are close to being at rest. These
authors also discovered that the
downflows immediately surrounding strong flux concentrations can reach supersonic
values at the solar surface, unlike in normal intergranular lanes, which display
only subsonic downflows in similar data. This provides observational confirmation that the additional
cooling of the gas in the immediate surroundings of magnetic concentrations
causes the surrounding gas to sink faster, as initially pointed out by
\citet{Deinzer1984}.

An early discovery of the observation of full Stokes~$V$ profiles was their
asymmetry. The blue and red wings of the profiles having different
amplitudes and areas \citep{Stenflo1984,Solanki1984,Solanki1985}, with the
asymmetry being in general larger in network features than in plage.
\citet{Stenflo1987} found the sign of the area asymmetry to change near the solar
limb. These results were confirmed by \citet{Martinez-Pillet1997} for plage
regions. The Stokes~$V$ area asummetry has
been interpreted in terms of (nearly) static gas inside flux tubes whose field
expands with height thus forming a canopy overlying the downflowing lanes of
the surrounding convective
cells \citep{Grossmann-Doerth1988,Solanki1989,Steiner1999,Shelyag2007}.
This geometry is now well-established and easily visible in high-resolution images
\citep{Berger2004,Pietarila2010}, see also the previous paragraph. This picture
could also explain the change in sign of the asymmetry close to the limb
\citep{Buente1993} and high resolution observations that show Stokes~$V$ to
be less asymmetric in the interiors of magnetic concentrations, but
strongly asymmetric near the edges, i.e. where the canopy is expected to be
located \citep{Rezaei2007,Martinez-Gonzalez2012}. \citet{Rezaei2007}
even find an opposite sign of the asymmetry inside magnetic features in the
examples considered by them. Such observations are a sign that magnetic flux
concentrations in plages and the network are becoming spatially
resolved and their internal structure probed.
Some Stokes $V$ profiles are anomalous in shape, i.e. having only a single lobe,
or three or four lobes. Such profiles are most common in the internetwork of
the quiet Sun, but are to some extent also found in active-region plage.
They can have different origins, with \citet{Sigwarth2001} proposing
network and plage fields as either a mixture of kG flux tubes and a sub-kG
field. This interpretation is supported by high resolution observations that
indicate weak opposite polarity fields surrounding kG magnetic features.

\citet{Martinez-Pillet1997} also found that plage fields contain many
so-called azimuth centers, i.e. regions where, starting
from the center of the feature, the magnetic azimuth points roughly isotropically in all
horizontal directions. They are thought to be signatures of intermediate-sized
magnetic features described by flux tubes expanding with height.
Inversions of data obtained by Hinode by \citet{Buehler2015}
have confirmed the presence of numerous azimuth centres in plage areas. These
authors also showed that the larger of these are associated with micropores.
Also visible are weak opposite polarity fields located just outside
the magnetic features in the deepest observable layers, underlying the canopy of
expanding fields in the upper photosphere \citep{Zayer1989,Narayan2011,Scharmer2013,Buehler2015}.

\subsection{Brightness of magnetic features in plage and the network}

The brightness of magnetic concentrations in plage and the network has been
a subject of intense study. The brightness gives insight into the transport
of energy within magnetic features, e.g., by the influx of radiation from the
surrounding granules \citep[e.g.][]{Spruit1976,Stenholm1977,Voegler2005,Holzreuter2012}.
Another strong motivation comes from the contribution
of facular and plage brightness to changes in the total and spectral solar
irradiance, the solar drivers of the Earth's climate
\citep[e.g.][]{Foukal2004,Domingo2009,Gray2010,Ermolli2013,Solanki2013} along
with other natural and man-made causes. Until recently the highest spatial
resolution was reached only, or at least the easiest, in broad-band images
taken in rapid bursts.

The variation over the solar cycle of solar surface area coverage by plage
also changes the intensity close to the solar limb (due to the enhanced
brightness of plage there) and hence produces variations in the apparent
solar radius (\citealt{Bruls2004}, cf. \citealt{Ulrich1995}).

The contrast of magnetic elements with respect to the quiet Sun increases
rapidly with height in the photosphere (e.g. see \citealt{Riethmuller2010}).
At the height of formation of the continuum in visible light the contrast
is low and can be of either sign. It is generally much stronger at
higher layers, sampled by, e.g., line cores
\citep[e.g.][]{Title1992,Stangl2005,Yeo2013}. The
continuum and line core contrasts also display a totally different
centre-to-limb behaviour. In the continuum the contrast is low at disk centre.
According to \citet{Topka1992,Topka1997} it is even negative for all types
of magnetic features. Negative contrasts of some types of small-scale magnetic features have
also been seen by \citet{Ortiz2002,Berger2007,Yeo2013}. The continuum contrast increases
toward the solar limb \citep{Domingo2005,Kobel2009}. Such a behaviour is also
observed for polar faculae \citep{Blanco-Rodriguez2007}. In many
observations, the continuum contrast increases from disk centre only
up to a given heliocentric angle ($\theta$), and then decreases towards the limb.
Thus \citet{Auffret1991} find that it peaks at around $\mu=\cos\theta\approx 0.35$,
while \citet{Ortiz2002} and \citet{Yeo2013} obtain that this angle depends on
the average flux density, i.e. probably the size of the magnetic feature,
which is in good agreement with results from flux-tube models.

In line cores, however, the largest contrasts are found near solar disk
centre, with the contrast decreasing towards the limb \citep{Yeo2013}.
\citet{Hirzberger2005} found no significant variation of the
contrast of faculae with heliocentric angle, from observations in G-band and
$587.5\pm1.5$~nm continuum. In the G-band this can be explained by the
mixture of lines and continuum, at 587.5 nm, it is more surprising.

The continuum contrast of magnetic elements is, on average, larger in the
network than in the plage \citep{Ortiz2006,Kobel2011,Romano2012}. This is
mainly caused by the larger average size of magnetic elements in these
regions, with magnetic features in plages being larger on average \citep{Grossmann-Doerth1994}. Larger
features are generally less bright around disk centre.

Correspondence of localised brightness enhancements to the magnetic flux
concentrations have been shown in MHD simulations
\citep{Voegler2005,Shelyag2007} as well as in the obsevations.
Observations at solar disc centre display a decrease in the contrast of
magnetic features with
field strength, either starting right from small field-strength values
\citep{Topka1992,Topka1997,Title1992}, or after an initial increase, i.e.
displaying a ``knee" in brightness
\citep{Lawrence1993,Stangl2005,Narayan2010,Schnerr2011,Kobel2011}.
Note, however, that observations made by \citet{Schnerr2011} with the very
high resolution SST (Swedish Solar Telescope; \citealt{Scharmer2003}) display an
increasing brightness right up to the largest field strengths that they
plot. This suggests that the spatial resolution does play a role in the
relationship between brightness and magnetic flux. \citet{Danilovic2013} studied
the effect of degrading simulated images to the spatial resolution of
Hinode/SP images (i.e. of limiting the spatial resolution; see also
\citealt{Rohrbein2011}). With the help of the appropriate point-spread-function
\citet{Danilovic2013} could reconcile the different rms contrasts obtained from
MHD simulations and observations. The apparent difference between the
contrast of magnetic features in observations and MHD simulations was
pointed out and explained by \citet{Rohrbein2011}. They showed that the application
of an appropriate point-spread-function turned the
monotonic relation between brightness and field strength found at the original
resolution of their MHD simulations into a relation with a maximum
at intermediate field strength.

A very high contrast is achieved by imaging within molecular bands, which have
the advantage that owing to the large density of lines, relatively broad filters
can be used, allowing a good signal-to-noise ratio to be reached within a
short exposure time. The most commonly used molecular band has been the
Fraunhofer G-band of CH molecular absorption
\citep{Muller1984,Muller1987,Berger1995,van-Ballegooijen1998,Berger2001,Nisenson2003,Langhans2004,Rouppe-van-der-Voort2005,Viticchie2010,Bodnarova2014}.
The bright points seen in this wavelength range are associated
with magnetic fields \citep{Berger2001,Bharti2006,Beck2007}, though not all
areas with a large magnetic flux can be necessarily associated with G-band
bright points \citep{Ishikawa2007}.
Although the CN-bandhead at 388~nm offers a larger density of lines and hence
produces stronger contrasts (e.g.
\citealt{Rutten2001,Berdyugina2003,Zakharov2005,Uitenbroek2007a}; but see
\citealt{Uitenbroek2006} for a deviating view), the G-band profits
from the often higher camera sensitivity and the larger number of solar
photons. The high contrast of the molecular bands comes from the dissociation
of the CH (respectively the CN) molecule in the higher temperature of
magnetic elements \citep{Steiner2001,Sanchez-Almeida2001}, which is well
reproduced by radiation-MHD simulations \citep{Schuessler2003,Shelyag2004}.

\begin{figure}[tbp]
      \centering
      \includegraphics[width=8cm, trim = 0 0 0 0, clip]{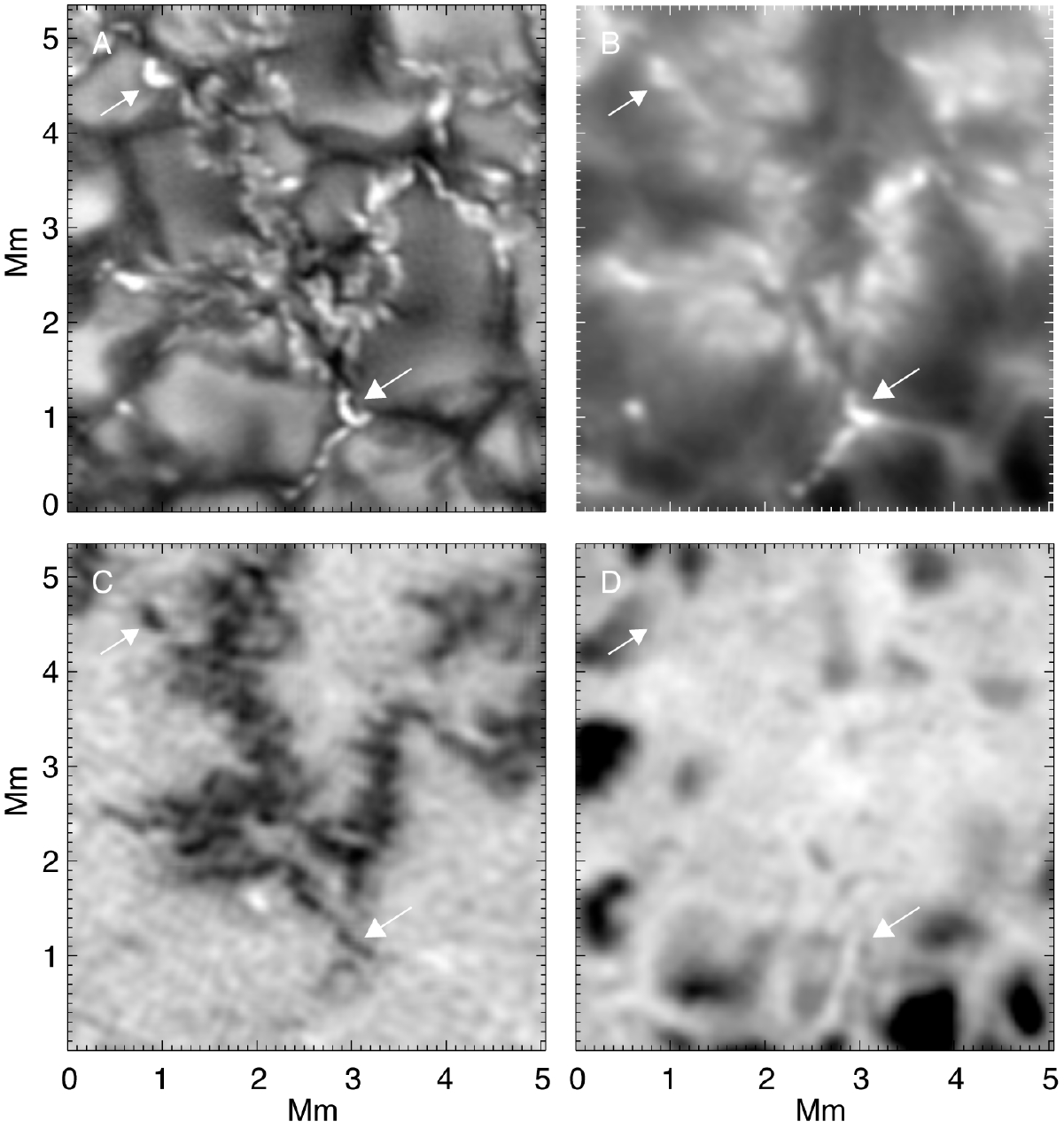}
      \caption{Brightness contrast in G-band (A) and Ca~{\sc ii}~H (B)
  passbands, from high-resolution observations with SST, as well as maps of magnetic
  flux density (C) and of line-of-sight velocity (D). The magnetic
  flux density and the line-of-sight velocity are linearly scaled, ranging
  from -1090 to 330~Mx/cm$^2$ and from -1.5 to 0.8 km/s, respectively.
  The arrows indicate loop-like emission structures over downflowing lanes studied by \citet{Berger2004} (see Sect.~\ref{subsect:obs_plage_network}).
  Reproduced with permission from \citet{Berger2004}, Astronomy \& Astrophysics, \copyright~ESO.
      }
      \label{fig:Berger2004}
\end{figure}

A small portion of an active region is imaged in the photospheric G-band and the low
chromospheric passband centred on the Ca~{\sc ii}~H line core in
Figs.~\ref{fig:Berger2004}A and \ref{fig:Berger2004}B, respectively (from
hight-resolution observations with SST; from \citealt{Berger2004}). They
clearly demonstrate the increasing contrast with height.
The Fe~{\sc i}~630.25~nm magnetogram is displayed in Fig.~\ref{fig:Berger2004}C,
for comparison with the location of the bright points. Also,
Fig.~\ref{fig:Berger2004}D presents the corresponding Dopplergram linearly
scaled between -1.5 and 0.8 km/s.

The main difference between plage and network magnetic features is their
size \citep[e.g.][]{Kobel2014}, i.e. the amount of magnetic flux
that they carry, as well as their motions. The brightest magnetic elements may
have, however, similar sizes in plage and the network \citep{Kobel2011}.
Some of the other differences are probably related to the size difference.
An example is their lifetimes, which are likely to be different in the network and
in plage, since larger magnetic features tend to live longer \citep{Meunier2009}.

A number of studies have found a correlation between the intrinsic field strength
of magnetic elements in network and plage and the magnetic filling factor, which
is a proxy of the amount of magnetic flux in a given spatial resolution
element
\citep{Stenflo1985,Schuessler1988,Zayer1990,Keller1990,Rueedi1992,Grossmann-Doerth1996,Martinez-Pillet1997,Sigwarth1999}.
In most of these studies the resolution element was considerably larger than
the individual magnetic features. The dependence may, again, be related to the
larger average size of magnetic features in regions with higher magnetic flux
density.

\subsection{Theoretical description of plage and network magnetic concentrations}

Solar magnetic flux concentrations have generally been described by a
flux-tube model (see, e.g.,
\citealt{Spruit1976,Spruit1977,Spruit1981b,Deinzer1984b}; see
\citealt{Solanki1993}, for a review of early work). It assumes that the
magnetic features can be described by a bundle of ordered field lines
that are enclosed by a topologically simple surface. In addition, the
features are assumed to be stationary and in force and energy balance
with their surroundings. Although reality is expected to be more complex
than this, for many purposes further simplifications are introduced. For
example, the shape of the cross-section of (vertical) flux tubes is
often considered to be round (i.e. having axial symmetry). This geometry
also builds on the near axial symmetry of the most regular
sunspots. Alternatively, sometimes flux sheets, with translational and
mirror symmetry, are considered (e.g., \citealt{Holzreuter2012}).  This
geometry describes better magnetic structures in regions with a larger
concentration of magnetic flux as in active-region plage. There the flux
has a tendency to fill the intergranular lanes, i.e. to concentrate into
``ribbon-like" structures \citep{Bushby2005}. In general, it is assumed
that a flux tube (used as a generic term also to describe features with
a non-circular cross-section) are surrounded by an electric current
layer where the magnetic field drops rapidly. The width of this layer is
estimated to be a few km \citep{Schuessler1986}.

A further and very commonly used simplification is the thin-tube approximation
\citep{Parker1955,Defouw1976,Pneuman1986,Ferriz-Mas1989}.
In its simplest form it reduces force balance to a simple horizontal and vertical balance of the
total pressure (composed of gas and magnetic pressure, with sometimes the
inclusion of turbulent pressure as well, which, however, is small compared to
the other two in the solar photosphere). The assumption is strictly valid only
for flux tubes whose radii are much smaller than the pressure scale height (which
in the photosphere is on the order to 100 km). Due to the presence of magnetic
pressure inside the tube, the gas pressure and hence also the density is
lower there, so that we see deeper layers (an effect known as the Wilson
depression in sunspots, where it can be directly observed; e.g. see
\citealt{Solanki2003a} for a review; also \citealt{Lites2004} for observations
at high spatial resolution).

\begin{figure}[tbp]
      \centering
      \includegraphics[width=6.5cm]{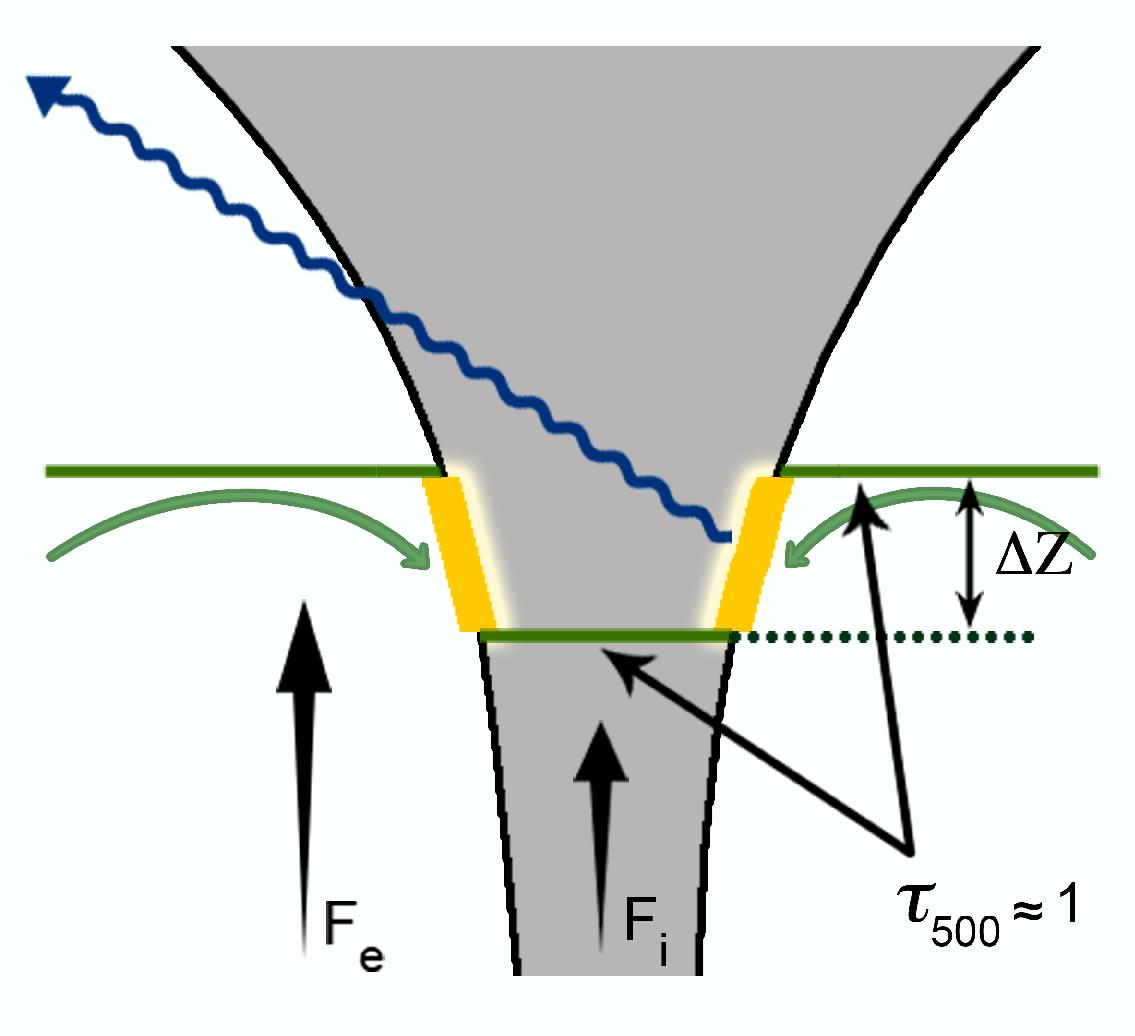}
      \caption{Sketch of the vertical cross-section of a thin flux tube
      (shaded, gray). The
      Wilson depression, $\Delta Z$, is the geometrical difference
      between the same optical depth ($\tau_{5000} = 1$) inside
      and outside of the flux tube (see main text). The radiating hot walls are
      indicated by yellow blocks. The green arrows illustrate
      the convection. The cartoon is not to scale. Inspired by a similar
      sketch by \citet{Schrijver2000}.
      }
      \label{fig:fluxtube}
\end{figure}

A sketch outlining the main points of a photospheric flux tube is given in
Fig.~\ref{fig:fluxtube}. It illustrates a typical, small-scale magnetic flux
concentration that appears bright. The evacuation inside the flux
tube lowers the optical depth inside the tube (Wilson depression) allowing
the radiation from the hotter gas inside the tube (compared to the external gas
at equal optical depth although not necessarily at equal geometric height) to escape and hence be observed. The gas inside the
tube is heated by radiation from the hot walls of the tube (yellow in the
sketch) through which the hot sub-surface gas in the surroundings radiates
excess energy, compared to the surroundings intergranular areas \citep{Spruit1976}.

Convective energy transport is strongly reduced or completely
quenched by the strong magnetic field inside magnetic flux tubes with kG
fields \citep{Gough1966,Ferriz-Mas1994}. Since
the vertical energy transport by radiation is comparatively inefficient
in the solar convection zone, for $\tau\gg1$ the vertical energy flux
density inside the flux tube, $F_{i}$, is much smaller than that in the
surroundings, $F_{e}$.  Consequently, most of the energy influx into the interior of
a flux tube comes through radiation flowing in from its walls (between the
$\tau=1$ levels inside and outside the flux tube), as
originally proposed by \citet{Spruit1976} and later demonstrated with the help of
numerical simulations by, e.g., \citet{Deinzer1984,Knoelker1988,Voegler2005},
cf.~\citet{Schrijver2000}.

As the gas pressure drops with height, so must the magnetic field in order to
maintain pressure balance. Magnetic flux conservation then causes the
magnetic field to expand with height, creating a canopy of field overlying
the convecting gas. Above a certain height the field of neighbouring magnetic features
reaches each other and if they are of the same polarity (which is common in
active region plage), they merge \citep{Spruit1983}. The rate of expansion of
the field is given mainly by the relative temperature inside and outside the
flux tube \citep{solanki1990}.
We note that the relative rates of expansion with height of large and small flux tubes are
similar \citep{Solanki1999}.

\begin{figure}[tbp]
      \centering
      \includegraphics[width=11cm]{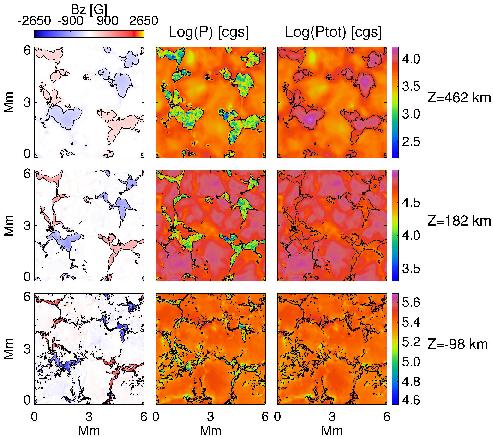}
      \caption{Stratification of the longitudinal component of the
      magnetic field ($B_{z}$; left column), gas pressure ($P$; middle column),
      and total pressure ($Ptot$; right column), with respect to the optical
      depth unity at 500~nm. The black contours include areas where
      $|B_{z}|>$500~G at -98~km, $|B_{z}|>$400 G at 182~km, and
      $|B_{z}|>$300~G at 462~km. Reproduced with permission from \citet{Yelles-Chaouche2009}, Astronomy \& Astrophysics, \copyright~ESO.
      }
      \label{fig:yelles2009}
\end{figure}

The MHD simulations clearly show that the magnetic flux concentrations are
neither mainly circular, nor always sheet-like (although examples of features
similar to both types can be found, see Fig.~\ref{fig:yelles2009} and
Fig.~\ref{fig:voegler2005}). Rather they constantly evolve, changing in
size and shape, and interact with each other. Although the shape of the cross-section
does not agree with that underlying the thin-tube approximation, the main
assumption, namely of horizontal
pressure balance, turns out to be satisfied if the second order
expansion of the thin-tube approximation, as derived by \citet{Pneuman1986} and
\citet{Ferriz-Mas1989}, is used (see \citealt{Yelles-Chaouche2009}).
Figure~\ref{fig:yelles2009} presents results of the MHD simulations by
\citet{Yelles-Chaouche2009}, where the vertical component of the magnetic
fields, gas pressure, and total pressure are plotted for three different
photospheric layers, with $z=0$ corresponding to the spatially averaged optical depth unity
at 500~nm.

\subsection{Comparison of theory and observations}

How does the model of the thin flux tube (or flux sheet) live up to observations?
One great success of this model has been that it could reproduce the
centre-to-limb variation of the brightness of magnetic features (or
equivalently of their contrast relative to the quiet Sun).

Now, thanks to the high-resolution
data available from sources such as Hinode \citep{Tsuneta2008}, the Swedish Solar
Telescope \citep{Scharmer2003} and the Sunrise balloon-borne observatory
\citep{Solanki2010,Barthol2011,Berkefeld2011,Gandorfer2011,Martinez-Pillet2011}
we can test with much higher fidelity to
what extent such descriptions live up to reality.

\begin{figure}[tph!]
      \centering
      \includegraphics[width=11cm, trim = 0 0 0 0, clip]{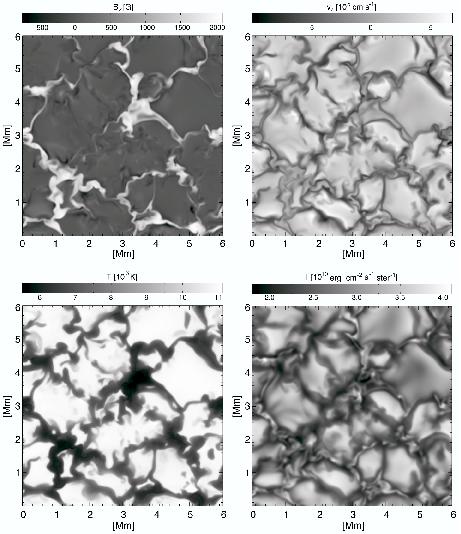}
      \caption{Maps of longitudinal magnetic field (upper left), line-of-sight
      velocity (upper right), temperature (bottom left), and intensity (bottom
      right) at the average geometrical height corresponding to optical depth
      unity (at 500 nm), from MHD simulations made with the MuRAM code.
      The strong magnetic field concentrations, in downflow areas, are
      surrounded by narrow lanes of weak opposite-polarity magnetic flux.
      Reproduced with permission from \citet{Voegler2005}, Astronomy \& Astrophysics, \copyright~ESO.
      }
      \label{fig:voegler2005}
\end{figure}

\begin{figure}[tph!]
      \centering
      \includegraphics[width=10cm, trim = 0 0 0 0, clip]{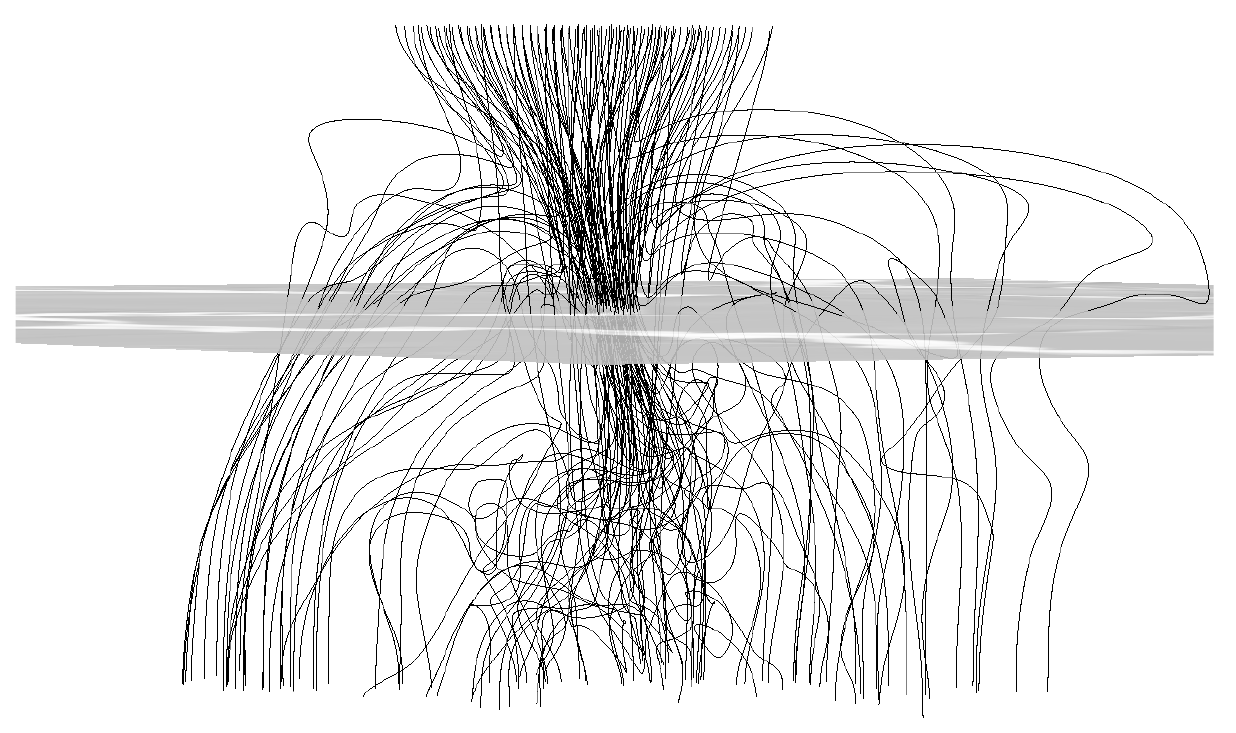}
      \caption{A side-view from MHD simulations, illustrating geometry of the field lines at magnetic flux concentrations (along a thin magnetic sheet) surrounded by weaker low-lying magnetic loops. The grey plane indicates the average geometrical height corresponding to $\tau_{5000} = 1$. Reproduced with permission from \citet{Voegler2005}, Astronomy \& Astrophysics, \copyright~ESO.
      }
      \label{fig:voegler2005_f2}
\end{figure}

Thus, the presence of weak opposite-polarity fields surrounding magnetic
flux concentrations has been established by MHD simulations, as can be seen
in Fig.~\ref{fig:voegler2005} taken from \cite{Voegler2005}, in good agreement with
the observations of such fields \citep{Zayer1989,Narayan2011,Scharmer2013,Buehler2015}.
The simulations reveal the geometry of the opposite polarity fields (which is not
accessible to the observations). These fields are due to field lines anchored within
the magnetic concentration that are turned over and pulled down by the surrounding downflows.
In general, they do not extend higher up in the atmosphere, but
rather most of these turn back at relatively low heights in the photosphere, just as in
the observations. Illustrated in Fig.~\ref{fig:voegler2005_f2} is a side-view of such structures from MHD simulations \citep{Voegler2005}, where concentrations of the field lines represent along-side-view of a thin magnetic sheet, surrounded by the low-lying field lines.

A remarkable success of MHD simulations has been the agreement of the
brightness (or brightness contrast) of simulated magnetic features with that
of observed ones. The centre-to-limb variation of the contrast at various
wavelengths has been modelled by \citet{Carlsson2004}, \citet{Keller2004},
and \citet{Penza2004}, cf. \citet{Steiner2005}.
The distribution of the brightness of magnetic bright points in
the G- and CN-bands has been successfully compared with the high-resolution
Sunrise observations by \citet{Riethmuller2010}. The RMS contrast in the same
images is dominated by granulation and was also reproduced by the same simulations
\citep{Hirzberger2010}. \citet{Riethmuller2014} went a full step
further and compared not just the brightness in broad wavelength bands, but
also a host of further parameters, such as the Stokes $V$ amplitude, the Doppler
shift, the line width, etc. and found a good agreement (with some slight
disagreement only in the width of the distribution of Doppler shift values)
between the values found in the quiet Sun by Sunrise and those from an MHD
simulation with an average vertical field of 30~G. The evolution of the fine-structure
of faculae has been investigated by, e.g., \citet{De-Pontieu2006}, who found,
using both high-resolution observations and MHD simulations, that faculae
change on a time-scale of minutes and that this change is mainly produced by
the evolution of the granules lying just behind them, whose hot sides are
better seen due to the evacuation of the faculae.

Disagreements in the observed and simulated rms contrasts found by, e.g.,
\citet{Uitenbroek2007}, are very likely due to the effect of scattered light
that is pervasive in ground-based data and even affects space-based observations
\citep{Danilovic2008}. The best observations with low scattered light, e.g.,
\citet{Hirzberger2010}, display a good agreement with simulated rms
contrasts. These results indicate the importance of the scattered light in
decreasing the contrast of high-resolution observations.

\subsection{Concentration of magnetic flux: convective collapse}

The formation of strong-field magnetic features was for a long time
studied mainly with theoretical methods. Two main mechanisms for the
concentration of magnetic flux were proposed. The first, flux expulsion,
represents a self-organized separation of magnetic field from
convection. Any vertically oriented magnetic field at the solar surface
is quickly transported to the intergranular lanes by the horizontal
flows within granules. This allows granular convection to continue
operating even in the presence of a significant amount of magnetic flux
(although the convection cells do change, becoming smaller, less
turbulent with smaller horizontal velocities, as well as more ordered
with a longer lifetime, e.g. \citealt{Title1989,Title1992,Narayan2010}).
The fields produced by this mechanism have roughly the same energy
density as the convective flows (i.e. the field is in equipartition with
the kinetic energy of the flow), which corresponds to roughly 200-400 G
at the solar surface.

The convective instability, acting on regions with such equipartition
fields can then intensify them to kG levels \citep{Parker1978, Webb1978,
Spruit1979a, Spruit1979b, Schuessler:1990, Takeuchi1997,
Grossmann-Doerth1998, Steiner:2003, Danilovic2010a, Hewitt2014}.  The
process can be idealized by the adiabatic instability of a static thin
magnetic flux tube with an equipartition magnetic field embedded in a
superadiabatically stratified medium. The instability leads to
accelerated downflow along the tube, which entails a reduction of the
internal gas pressure.  The excess pressure in the surroundings then
compresses and enhances the field until the magnetic pressure becomes
sufficiently large to compensate for the loss of gas pressure and a new
equilibrium is reached (see \citealt{Spruit1979b} for a discussion of
the non-linear development of the instability).  The final state needs
not to be a stationary one, with \citet{Hasan1985} finding an
oscillatory end state in his 1D model.

In reality, a static equilibrium with equipartition field will most
probably never be realized. Instead, the magnetic flux advected by the
horizontal granular flow accumulates in intergranular downflow
regions. Once the horizontal flow is suppressed by the growing field
strength, the downflow is no longer supplied by these flows, yet
continues owing to the superadiabatic stratification and the ongoing
surface cooling by radiation. This leads to partial evacuation and
growth of the field strength \citep{Schuessler:1990}.

The 3D radiation MHD simulations of \citet{Danilovic2010a} and
\citet{Hewitt2014} followed all steps of the process from the gathering
of flux at the intergranular lanes, followed by an evacuation produced
by a downflow and the associated strengthening of the field that then
quenches the downflow. The field can get weakened again, in some cases
possibly due to the following upflows \citep{Grossmann-Doerth1998,
Danilovic2010a}, but at least in some cases without the need of an
upflow \citep{Hewitt2014}.

If magnetic features are below a certain size (or below a certain amount of
magnetic flux) then the radiative exchange of energy with the surroundings
becomes strong enough to reduce or to completely compensate for the cooling of
the downflowing material, so that the collapse does not occur. Hence according to theory
the features with the smallest magnetic flux should possess weak fields
\citep{Venkatakrishnan1986,Solanki1996,Grossmann-Doerth1998}.

Observations have in general been lagging behind the theory, with the first
suggestion of a correlation between the magnetic flux of a feature and its intrinsic field strength being found by Solanki et al. (1996). The observed strengthening of the field was first interpreted as a convective collapse
by \citet{Bellot-Rubio2001}. The increase in the field strength in this case
was from 400~G to only 600~G. Larger increases have been obtained by the more
recent investigations by \citet{Shimizu2008,Nagata2008,Fischer2009,Narayan2011,Requerey2014,Utz2014}.
The final two
papers study the evolution of small magnetic elements in the quiet Sun. They
find that, at least for these features with relatively small amounts of flux,
the strong-field state does not last very long, with the field strength
fluctuating in an almost periodic fashion (cf. \citealt{Martinez-Gonzalez2011}).
These studies suggest that the evolution of the field strength of the larger and stronger
magnetic concentrations making up the bulk of active region plage would be a promising topic of study.

\section{Recent discoveries on internetwork magnetic fields}
\label{sec:internetwork}

\begin{figure}
\begin{center}
\includegraphics[width=16cm,trim = 45 400 0 0,clip]{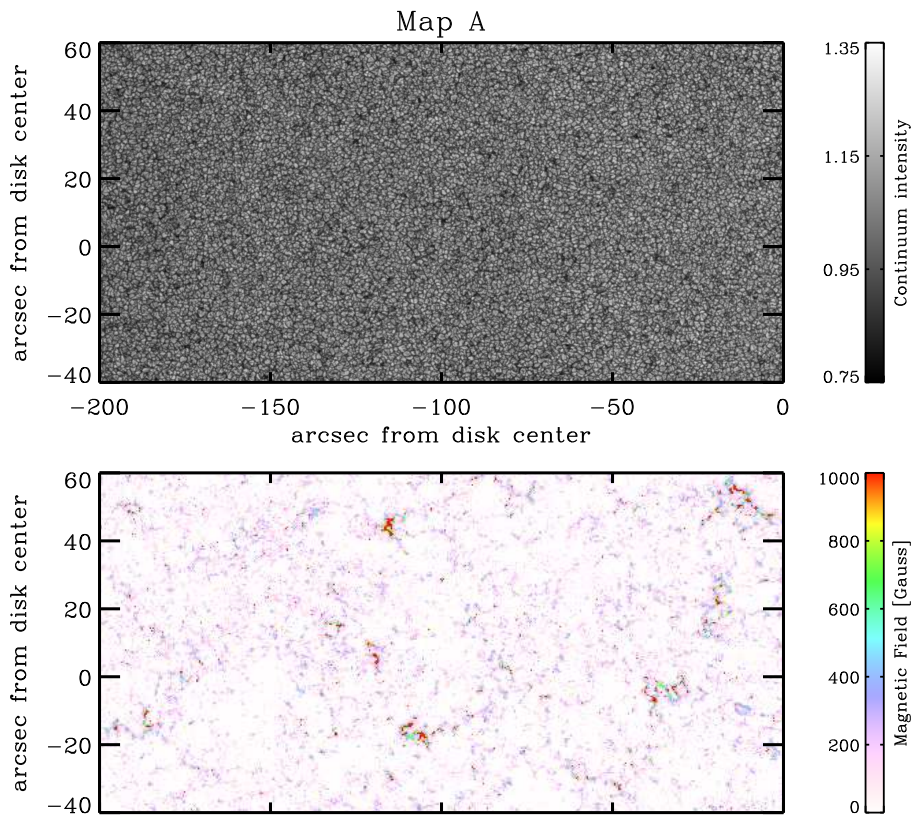}
\end{center}
\caption{Continuum intensity normalized to its average value over the
         entire field of view (top) and magnetic field strength (bottom)
         inferred from Hinode observations of the quiet Sun very close
         to disk center \citep{borrero2011pdf}. The network on the lower
         map can been identified in green/orange/red colors, whereas the
         internetwork can be seen as the regions in pink/white
         colors. The latter covers at least 90 \% of the total observed
         area. This particular map has been employed by many authors to
         study the angular distribution of the magnetic field in the
         internetwork (see Sect.~\ref{sec:angulardistribution}).
         Reproduced with permission from Astronomy \& Astrophysics,
         \textcopyright ESO}
\label{figure:hinodemap} 
\end{figure}

Despite being much more difficult to detect and measure than magnetic
fields in sunspots, plage and network regions, internetwork (IN in what
follows) magnetic fields are of great importance for several reasons.
Firstly, there is the possibility that a considerable amount of magnetic
flux in the Sun's quietest regions (i.e., the internetwork) is provided
by a turbulent small-scale dynamo. Since numerical simulations cannot be
carried out for the extreme solar values of Reynolds numbers and
magnetic Prandtl number (see Sect.~\ref{sec:ssd}), observations are
indispensable to test the validity of the computational results and
possibly prove the existence of small-scale dynamo action.  Secondly,
the relatively weak internetwork magnetic fields (see
Fig.~\ref{figure:hinodemap}) interact in a highly dynamical fashion with
convective motions in the solar surface layers. This gives rise to
highly interesting small-scale phenomena and convoluted magnetic
topologies \citep{amari2015heating}.  Thirdly, if IN magnetic fields 
are produced by a small-scale dynamo and considering that at least 
90 \% of the solar surface is covered by the IN\footnote{This number
should be considered as a lower threshold as sometimes it is not easy to
distinguish between strong isolated IN fields and the network.}, 
this would mean that most of the unsigned photospheric magnetic flux 
does not vary over the 11-year magnetic
cycle. This could have important consequences for our understanding of
the solar irradiance variations, which contribute to terrestrial climate
variations \citep{solanki2013irradiance, yeo2014irradiance}.

The main question is whether IN magnetic fields do actually arise from a
turbulent small-scale dynamo or are leftovers of the magnetic fields
from decaying active regions, being spread out over the solar surface by
advection due to supergranulation, meridional flow, and differential
rotation. Therefore, characterizing the internetwork magnetic fields
can be considered as one of the most pressing issues in solar physics.

\subsection{Angular distribution of internetwork fields}
\label{sec:angulardistribution}

A small-scale dynamo operating in a stratified medium provides an
isotropic magnetic field at scales smaller than the pressure scale
height. Numerical simulations confirm this expectation in the
sub-surface layers, but predict a strong anisotropy in favor of
horizontal fields in the photosphere (i.e., in convectively stable
layers), both for the small-scale dynamo \citep{Schuessler:Voegler:2008,
Rempel:2014} and for magnetoconvection with a mean vertical field
\citep{steiner2009simul, steiner2012pdf}.  Consequently, observationally
confirming or ruling out such a distribution has been a major field of
research over the past decade. Unfortunately, inferring the magnetic
field in the internetwork is extremely challenging owing to their weak
observational signatures (i.e. polarization signals barely above the
noise). Not surprisingly, these difficulties have led to seemingly
contradictory results. The situation is further complicated by the
varied analysis techniques employed and the different kinds of data
analyzed. In the following we present a summary of observational results
organized in terms of the inferred angular distribution for IN magnetic
fields.

\subsubsection{Preference for horizontal fields}
\label{subsec:horizontal}

\cite{orozco2007pdf} and \cite{lites2008pdf} analyzed low-noise ($\sigma
\approx 3\times 10^{-4}$)\footnote{The noise level $\sigma$ is given in
units of the continuum intensity spatially averaged over the quiet Sun.} 
data from the Fe {\sc I} line pair at 630~nm recorded at disk center with 
the spectropolarimeter on-board the Hinode satellite. Employing 
inversion techniques and different calibration curves they concluded 
that magnetic fields in the
internetwork have a clear preference for the horizontal (i.e. parallel
to the solar surface) direction, with the horizontal magnetic field
being much stronger (in a spatially average sense) than the vertical one. However, \cite{stenflo2010pdf}
and \cite{borrero2011pdf,borrero2012pdf} later argued that this result
is likely due to photon noise. Indeed, if we consider a weak magnetic
field whose projection along the observer's line of sight is the same as
the projection on the plane perpendicular to the line of sight,
$B_\parallel = B_\perp$, the 630~nm lines provide much smaller (about 10
times weaker) signal in linear polarization (Stokes $Q$ and $U$) than in
circular polarization (Stokes $V$). Consequently, the same amount of
photon noise in the observed circular and linear polarizations is
interpreted as a magnetic field with $B_\perp \gg B_\parallel$, i.e., a
preference for horizontal orientation.

To circumvent this issue, \cite{bellot2012pdf} analyzed data from the
same instrument, recorded also at disk center, but averaged in time over
24 minutes. This allowed them to decrease the photon noise to $\sigma
\approx 7\times 10^{-5}$ such that almost 70 \% of the internetwork
displayed linear polarization signals clearly above the noise. The
magnetic fields inferred from the analysis of these regions again showed
a clear preference for the horizontal direction.  However, the long
integration times needed to decrease the noise might introduce
additional effects (such as the cancellation of signals from opposite
polarities and averaging over different structures) that need to be
carefully addressed.

\subsubsection{Preference for vertical fields}
\label{subsec:vertical}

\cite{ishikawa2011pdf} studied, using also calibration curves, the very
same data set as \cite{orozco2007pdf} and \cite{lites2008pdf}.  However,
they concluded that the vertical magnetic field in the internetwork is
almost as strong as the horizontal field. This would indicate a larger
contribution from vertical (perpendicular to the solar surface) magnetic
fields than in the aforementioned works.  However, we note that the low
value of the horizontal magnetic flux in \cite{ishikawa2011pdf} follows
from their procedure to ascribe a zero value to the perpendicular
component of the magnetic field whenever the linear polarization was
deemed to be at, or below, the noise level ($B_\perp = 0\;$ if the
line-integrated unsigned linear polarization $\;L_{\rm tot} \leq 1.5
\times 10^{-4}$). This approach has the problem of potentially deriving
vertical magnetic fields from signals that are actually produced by
strongly inclined fields. This happens because the most that can be said
about signals with $L_{\rm tot} \leq 1.5 \times 10^{-4}$ is that
$B_\perp$ is below some threshold value $B_{\rm thr,\perp}$. The problem
is that if the value of this threhold is such that $B_{\rm thr} \gg
B_\parallel$ then this approach will heavily underestimate the
contribution of $B_\perp$ to the total magnetic field.

Mostly vertical fields have also been reported by
\cite{jafarzadeh2014pdf} using geometrical considerations in
internetwork magnetic bright points seen at different photospheric 
layers. Unfortunately this method is not applicable to the majority
of the internetwork (i.e., outside magnetic bright points).

\subsubsection{Isotropic and quasi-isotropic distribution}
\label{subsec:isotropic}

\cite{asensio2009pdf} also analyzed spectropolarimetric data at disk
center from Hinode (see Fig.~\ref{figure:hinodemap}), but applied a
Bayesian technique to observations with a larger photon noise ($\sigma
\approx 10^{-3}$) than all previously cited works. This resulted in an
inferred angular distribution of the magnetic field in the internetwork
that is close to isotropic. \cite{asensio2014pdf} reached the same
conclusion, although they could not exclude the existence of a
non-negligible contribution from highly inclined magnetic fields.

An isotropic distribution was also favored by
\cite{martinez2008pdf}. This work employed two spectral lines in the
infrared (Fe {\sc I} at 1565~nm) that are less affected by photon noise in the
linear polarization, thus leading to more reliable inferences of
$B_\perp$. Unlike most of the previous works, spectropolarimetric
observations of the internetwork at various heliocentric viewing
angles were considered. The resulting distribution of observed
polarization signals (not of inferred magnetic fields) was found to be
nearly independent of the viewing angle, suggesting an isotropic
distribution of the IN fields. However, the authors emphasized that the
observed distributions of polarimetric signals lacked a
sufficiently large number of occurrences for signals above
$4 \times 10^{-3}$ (in units of the continuum intensity) to be considered 
statistically significant.

\citet{borrero2013pdf} also studied the distributions of polarimetric
signals and tried to obtain more reliable statistics of
spectropolarimetric observations at different heliocentric angles. To
that end, they analysed a large data set from the Hinode satellite with
a noise level of $\sigma \approx 3\times 10^{-4}$. They found that the
distribution of signals does in fact depend on the heliocentric angle,
indicating a non-isotropic distribution of magnetic fields in the
IN. 

\subsubsection{Bi-modal distribution}
\label{subsec:bimodal}

The same data (see Fig.~\ref{figure:hinodemap}) as \cite{asensio2009pdf}
and \cite{asensio2014pdf} was analyzed by \cite{stenflo2010pdf}, who
employed a line-ratio technique aided by several calibration curves in
order to to disentangle thermodynamic and magnetic effects in the
observed signals. He found evidence of two distinct populations of
polarimetric signals in the internetwork. A population with a slope of
the line-ratio\footnote{The line-ratio technique measures the ratio {\it
s} of the amplitudes in the circular polarization signals in two
different spectral lines. Under the assumption that both lines possees
the same opacity and provided that the magnetic field is weak, then {\it
s} is equal to the quotient of their Land\'e factors.} of $s \approx
1.15$ was attributed to strong (i.e. kilogauss) and mostly vertical
magnetic fields, which remained unresolved at the spatial resolution of
Hinode observations ($\approx 0.3$"). The second population
of signals with a slope of $s \approx 1.66$ was ascribed to weak and
isotropically distributed magnetic fields.  These findings were further
investigated by \cite{steiner2012pdf}, who analyzed polarimetric signals
synthesized by solving the radiative transfer equation with input from
three-dimensional MHD simulations.  They also found the two populations
with the same slopes as reported by \cite{stenflo2010pdf}.  However, the
two populations resulted from magnetic fields that, while all being
weak, increase or decrease with height in the photosphere. Regardless of
the physical realism of these simulations, the results by
\cite{steiner2012pdf} clearly point towards an uniqueness problem in the
interpretation of the observed signals via the line-ratio technique.

\subsubsection{Further results}
\label{subsec:furtherresults}

It has now become clear that data recorded at disk center are not
sufficient to uniquely determine the angular distribution of the
magnetic field in the internetwork. Indeed, the most recent attempts
have focused on observations taken at several heliocentric angles, but a
consensus has still not been reached. Using spectropolarimetric data
from Hinode, \cite{orozco2012pdf} concluded that the distribution of the
inclination of the magnetic field has a preference for the perpendicular
direction (with respect to the line of sight) at all heliocentric angles
studied ($\mu=1.0,0.5,0.23,0.1$). This result can be explained by
magnetic fields that, being mostly horizontal at
disk center, become gradually perpendicular to the solar surface for
smaller $\mu$ values. We emphasize, however, that these results are
possibly affected by the same bias as those of \cite{orozco2007pdf} (see
Sec.~\ref{subsec:vertical}). \cite{borrero2013pdf} also interpreted
Hinode data in terms of IN magnetic fields that become gradually more
horizontal for decreasing $\mu$. This should not be considered a general
trend because these authors employed only data at $\mu=1.0$ and
$\mu=0.7$.  In contrast, \cite{stenflo2013pdf} concluded from
spectropolarimetric observations of the Fe {\sc I} lines at 525 nm taken with
the ZIMPOL instrument that magnetic fields at $\mu=0.5$ are more
perpendicular to the solar surface (i.e., more vertical) than at
$\mu=0.1$.

Several effects can be invoked to explain why the angular distribution of IN
magnetic fields changes with the heliocentric angle $\mu$. For instance we
could take into account that the height of formation of the
spectral lines changes with $\mu$ because the continuum optical depth
veries with the viewing angle: $\tau_c=f(\mu)$. This would readily
introduce a $\mu$-dependence in the angular distribution of IN magnetic
fields if, as suggested by numerical simulations
\citep{Schuessler:Voegler:2008, steiner2009simul, steiner2012pdf,
Rempel:2014}, the angular distribution of the magnetic field changes with height
in the photosphere. Alternatively, if $\tau_c$ is a slowly varying
function of $\mu$ or if the angular distribution of the magnetic field does not vary with height on
the photosphere, one could ascribe the $\mu$-variations of the angular
distribution to, for instance, an instrinsic difference between IN polar fields 
and those at the equator. In addition, it should be considered
that the spatial resolution of the observations depends on $\mu$, because,
due to foreshortening, and given angular resolution element covers a larger
piece of the solar surgace near the limb than at disk centre. While 
in actuality all these effects are likely to play a role, it is not clear 
how to model them when interpreting the observations. A way to study 
which of the first two aforementioned effects is more important would be 
by means of high signal-to-noise spectropolarimetric observations along
 the equator and along the central meridian.

\subsubsection{Future prospects}
\label{subsec:futureprospects}

In spite of many efforts, we are currently no closer to accurately determining the 
angular distribution of the magnetic field in the internetwork than we were a
decade ago. This results from a combination of
the effects of photon noise and signal selection, as well as from the
different techniques employed to infer the magnetic field from
spectropolarimetric signals produced by the Zeeman effect.  Advances on
this front could be achieved by means of spectropolarimetric
observations in the infrared. In particular, the Fe {\sc I} line pair at
1565~nm is much less affected by noise in the linear polarization (see
Sect.~\ref{subsec:isotropic}). Moreover, the continuum of these spectral
lines is formed deeper (about 70-100 km deeper depending on the opacity
sources included in the calculation of the absorption coefficient) than
in the commonly used Fe {\sc I} line pair at 630~nm. Therefore, these spectral
lines are best suited to study the angular distribution of the magnetic field in
the lower layers of the photosphere (see
Sect.~\ref{sec:angulardistribution}). The only drawback comes from the
lower spatial resolution in the infrared, but this will be alleviated in
the near future with observations using large-aperture telescopes such as
GREGOR and DKIST, thanks to their large primary mirrors and hence smaller 
diffraction limits. The former one, with a 1.5-meter aperture, is already 
obtaining spectropolarimetric observations in these spectral lines with 
a spatial resolution comparable to that of the Hinode telescope in the 
visible \citep{collados2012gregor}.

Spectral lines whose spectropolarimetric signals are produced by the
Hanle effect are also worth considering for future observations and
analysis. Indeed, in the saturated Hanle regime the linear polarization
signals depend only on the orientation of the magnetic field, but not on
its magnitude \citep{carlin2015hanle}. This removes an important source
of uncertainty. In addition, Hanle signals are less affected by cancellation
effects if the magnetic field is unresolved. This feature of the Hanle
effect is particularly relevant for investigations of the magnetic field
produced by a small-scale turbulent dynamo. It remains to be seen,
however, how accurately our theoretical understanding of the Hanle effect and
its intricate interactions with the thermodynamics and kinematics of
the plasma \citep{carlin2012hanle} allow us to determine the properties
of IN magnetic fields.

\subsection{Solar cycle variations}
\label{sec:solarcyclevariations}

The presence of a small-scale turbulent dynamo that operates
independently of the large-scale dynamo responsible for the 11-year
activity cycle would provide a time-independent background level of
unsigned magnetic flux in the quiet Sun.

\begin{figure*}
\includegraphics[width=6cm,trim = 0 0 0 0,clip]{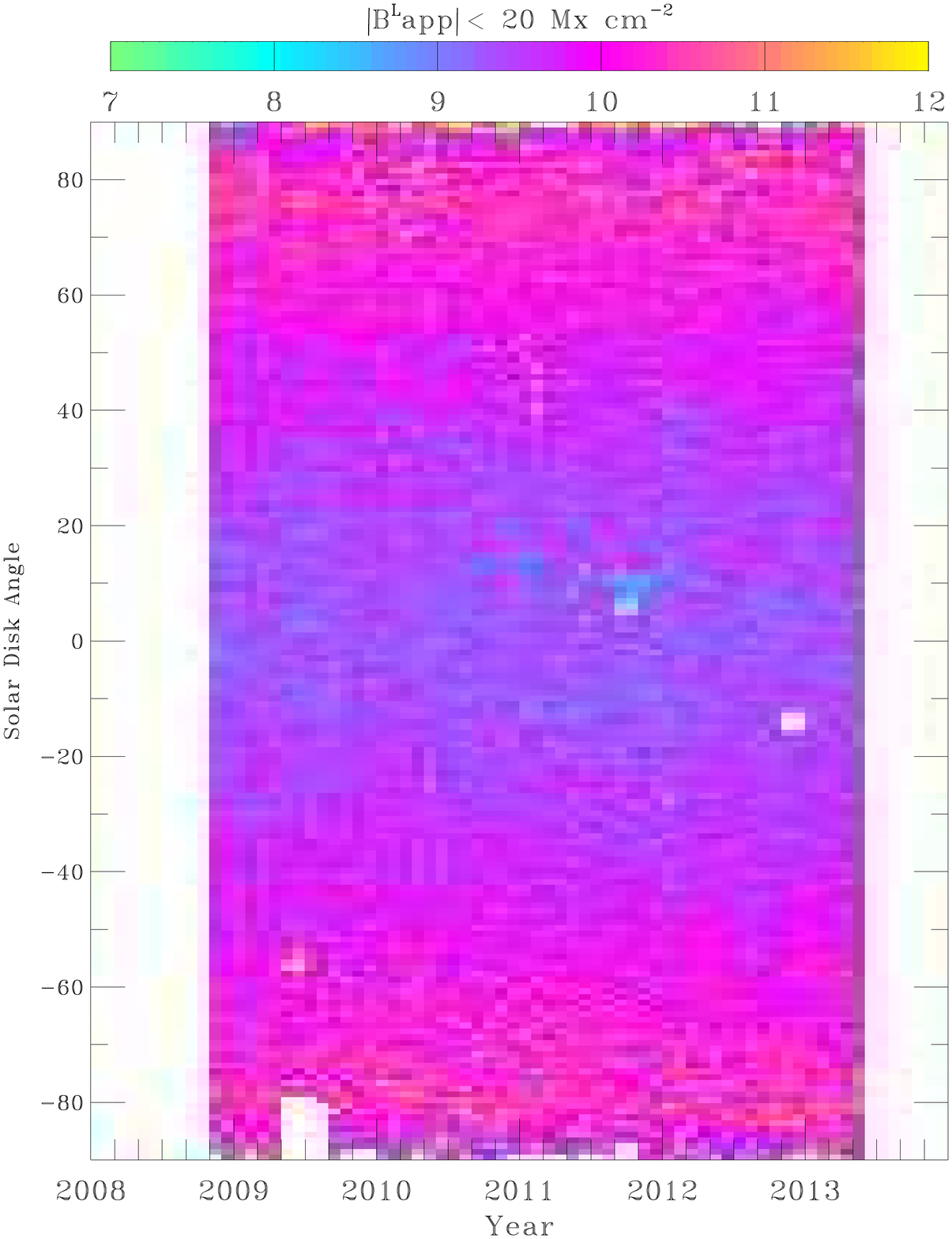}
\includegraphics[width=6cm,trim = 0 0 0 0,clip]{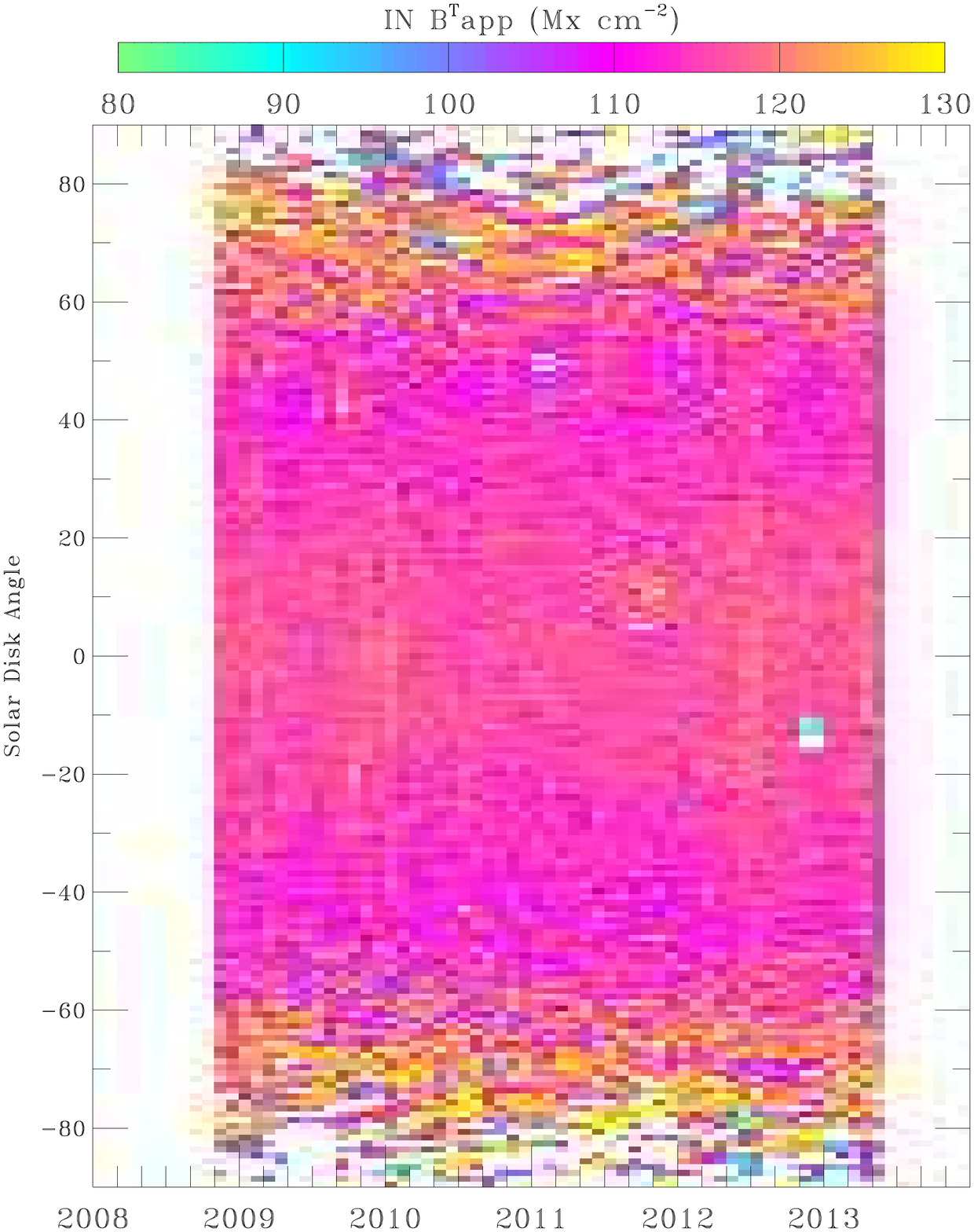}
\caption{Unsigned longitudinal component (left panel) and transversal
  component (right panel) of the magnetic field in the internetwork as a
  function of time (abcissa) and solar latitude (ordinate, from
  $-90^\circ$ to $+90^\circ$ with tickmarks at intervals of
  10$^{\circ}$). The time coverage spans from 2009 (close to
  solar minimum) until 2013 (close to solar maximum). Figure adapted
  from \cite{lites2014cycle}, reproduced with permissions from the
  Oxford University Press on behalf of the Astronomical Society of
  Japan.}
\label{figure:lites_cycle_blos} 
\end{figure*}

From ground-based observations \cite{almeida2003cycle} concluded that
there is no correlation between the strength of the magnetic field in the
internetwork and the solar cycle. However, to study this particular
aspect of the magnetic fields in the internetwork, it is necessary to
employ high-quality long-term observations of the polarization signals
in spectral lines. This is the case for the synoptic observations
carried out by Hinode's spectropolarimeter. \citet{buehler2013cycle,
jin2015cycle1, jin2015cycle2} analysed these data and concluded that
during the period between 2006-2015 the magnetic field in IN regions at
disk center did not vary significantly.  This result was later extended
by \cite{lites2014cycle} (see Fig.~\ref{figure:lites_cycle_blos}) to 
cover both low and mid latitudes ($\mu \ge
0.5$). While these studies were based on the Zeeman effect in Fe {\sc I}
lines at 630~nm, \cite{kleint2010cycle} and \cite{bianda2014cycle}
analysed spectropolarimetric data from the ZIMPOL instrument in the SrI
line at 460 nm and in molecular C$_2$ lines at 514~nm. The polarization
signals of the recorded spectral lines are produced by the Hanle effect,
which is much less affected by cancellation than the Zeeman effect for
magnetic fields of opposite polarities. These authors concluded
that for the period between 2007 to 2009 (solar minimum) the magnetic
field strength in the IN remained the same. However, they report a
possible variation with respect to earlier measurements obtained in 2000
(solar maximum).

\subsection{Small-scale events}
\label{sec:smallscaleevents}

While most polarimetric signals in the internetwork are at or below the
noise level of the observations (Sect.~\ref{sec:angulardistribution}),
some features show particularly strong signals. Detailed analysis of 
these revealed a variety of dynamic 
magnetic phenomena taking place in the internetwork. This is
not surprising because the magnetic field in the internetwork is not
strong enough to dominate over convective motions. Instead, these
motions twist and bend the field lines into complicated patterns. 
This leads to large variations of the magnetic, kinematic and
thermodynamic properties on small scales, which manifest
themselves in complex spectropolarimetric signals featuring
large asymmetries and additional components or lobes.

Since the observational signals associated with such small-scale
phenomena are very dynamic and short-lived, high-cadence instruments
attached to seeing-free telescopes (Hinode, Sunrise) or large-aperture
telescopes at the ground (SST, GREGOR), are required to characterize
them. In the following we summarize some of the recent results.

\subsubsection{Emergence of magnetic loops}
\label{subsuc:emergenceloop}

The emergence of magnetic $\Omega$-loops has been studied in detail, by
\cite{centeno2007emergence} and \cite{martinez2009emergence} using
Hinode data. Such loops are observed as single patches of linear
polarization (indicating horizontal magnetic fields) that are seen prior
to the appearance of two patches of opposite circular polarization
(indicating magnetic fields of opposite polarities) at the endpoints of
the elongated horizontal field patch. As it evolves in time, the
horizontal magnetic field disappears and the footpoints of opposite
polarities separate. While these observations are, in principle,
compatible with both an emerging $\Omega$-loop and a submerging inverted
$\Omega$-loop, the former interpretation is preferred whenever the
initial horizontal field usually is found in upflow regions (i.e.,
granules). Those appearing in downflow regions are therefore associated to
inverted $\Omega$-loops \citep{pietarila2011loop}

The lifetime of a few minutes and size of 1-2"
of the features are very similar to corresponding quantities of
granules. From observations of the Mg {\sc I} 517~nm and Ca {\sc II} 396~nm spectral
lines (formed in the mid and lower chromosphere, respectively)
\cite{martinez2010emergence} concluded that many of the loops
reach the lower chromosphere, carrying magnetic energy at a rate of
$\approx 2\times 10^6$ erg cm$^{-2}$ s$^{-1}$.

However, not all horizontal fields observed in the internetwork are part
of an emergence process. \cite{Danilovic:etal:2010b} analysed
spectropolarimetric observations of the Fe {\sc I} line at 525~nm with the IMaX
instrument on-board Sunrise. They studied a large number of internetwork
horizontal field patches and concluded that, although they 
tend to appear at the boundary of granules, most of them are caught at 
some point in their evolution in an intergranular lane.
This makes it difficult to uniquely ascribe them to a emergence of 
submergence event. These authors also found that horizontal IN fields do 
not possess a typical life-time or size.

\subsubsection{Siphon-flows along magnetic loops}
\label{subsec:siphonflow}

\begin{figure}
\begin{center}
\includegraphics[width=12cm]{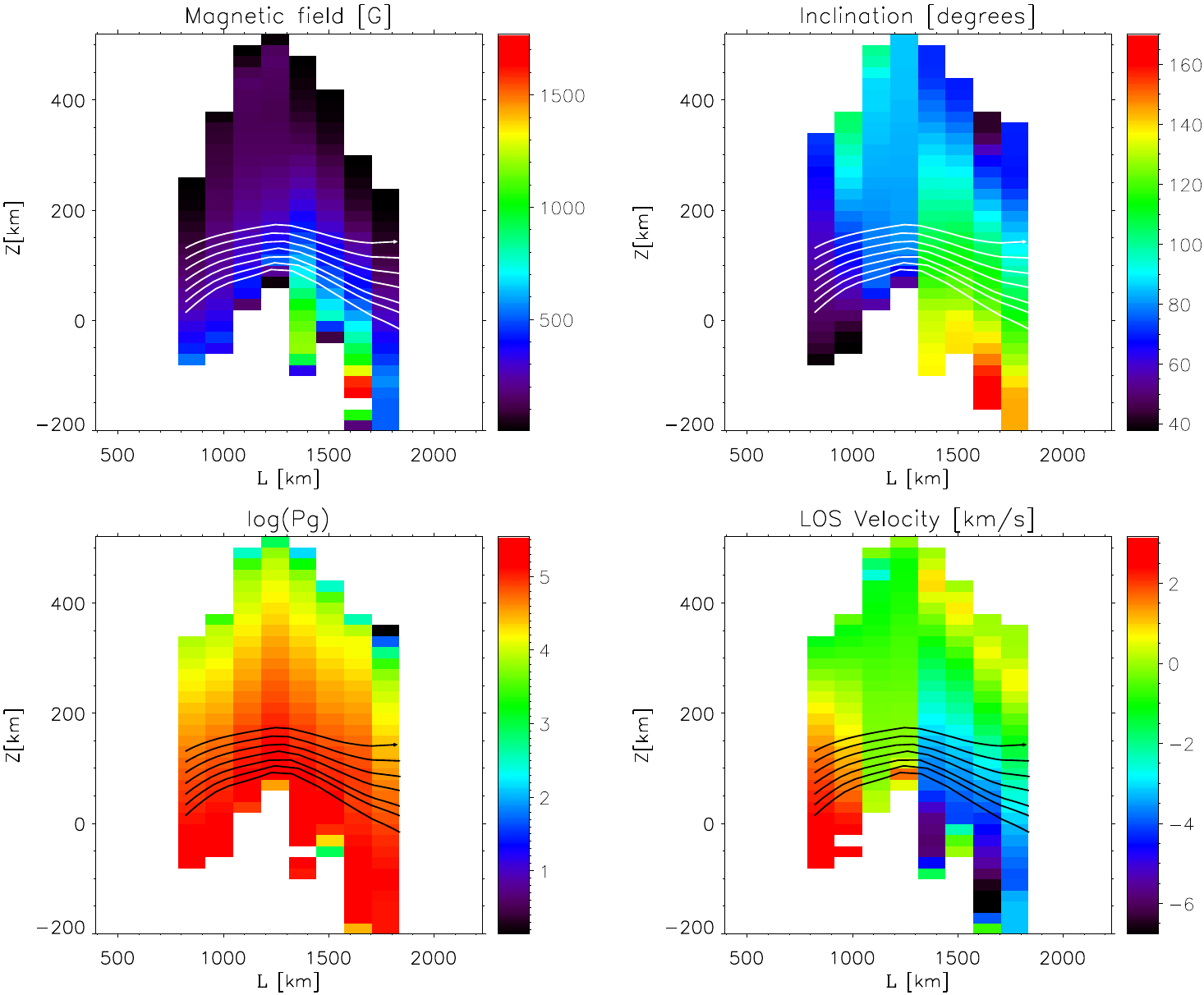}
\end{center}
\caption{Side-view of a magnetic $\Omega$-loop seen in Hinode/SP data \citep{quintero2014siphon}: 
magnetic field strength (upper-left), magnetic field inclination with respect to the vertical
direction (upper-right), gas pressure (lower-left), line-of-sight velocity 
(lower-right). The magnetic field at the left and right footpoints of the loop points 
upwards and downwards, respectively. On the other hand, the velocity 
points downwards and upwards at the left and right footpoints, respectively, 
with the latter reaching values up to $-6$ km s$^{-1}$ (supersonic). Reproduced with 
permission from Astronomy \& Astrophysics, \textcopyright ESO.}
\label{figure:siphon_loop} 
\end{figure}

Highly asymmetric, single-lobed Stokes $V$ profiles are often observed
at the footpoints of the magnetic loops discussed in the previous
subsection \citep{sainz2012emergence}. \cite{viticchie2012emergence}
interpreted them as being produced by a siphon flow along the magnetic
loop \citep{meyer1968siphon_1,meyer1968siphon_2}. This conclusion is
supported by \cite{quintero2014siphon}, who also found that the flow has
the opposite direction at the two footpoints and can become supersonic in
between (see Figure ~\ref{figure:siphon_loop}). Supersonic siphon flows 
had been theoretically predicted by \cite{thomas1990siphon} and already observationally 
detected by \cite{ruedi1992siphon} and \cite{degenhardt1993siphon}.

\subsubsection{Supersonic upflows}
\label{subsec:reconnection}

\cite{borrero2010super} found highly blue-shifted Stokes $V$ signals in
Sunrise/IMaX data, appearing close to, but not exactly at, locations
with large linear polarization signals (i.e., horizontal magnetic
fields). They ascribed these signals to supersonic magnetic upflows
occurring at the centers or edges of granules. According to
\cite{quintero2013super} the detection of the linear polarization signal
precedes the neighbouring supersonic upflow (highly blue-shifted
circular polarization). \cite{rubio2015super} studied many such
events at different positions on the solar disk and found that their
properties do not vary significantly with the heliocentric angle.

\cite{borrero2013super} suggested that the supersonic upflows were
caused by the reconnection of an emerging $\Omega$-loop with an ambient
field of opposite polarity \citep{parker1972reconnection,
parker1973reconnection}.  However, this
particular interpretation has been questioned by
\cite{danilovic2015bubbleup} who, analyzing synthetic
spectropolarimetric data obtained from MHD simulations, found that the
same observational features can also be reproduced in the absence of
reconnection. Instead they propose a magnetic topology that closely
resembles that of a plasma bubble filled with a strong unipolar magnetic
field that rises through the photosphere. Structures with very similar
magnetic topologies but moving downwards had previously been reported by
\cite{quintero2014bubbledown}.

\subsubsection{Magnetic elements in the internetwork}
\label{subsec:magneticelements}

Magnetic flux concentrations sufficiently large to appear in the
form of bright points are also present in the internetwork but, unlike
in the network, they appear isolated and are less abundant: 2.2 bright
points per Mm$^2$ in the network versus 0.85 in the internetwork
\citep{almeida2010magneticlements}. Note that, in addition, smaller
kG flux concentrations may exist that are not resolved by the observations.

The properties of the internetwork elements are rather diverse.  Some of
them are observed as mostly horizontal fields with a lifetime of 1-10
minutes and typical sizes of several arcseconds. These \emph{transient
horizontal fields} (THFs) were previously detected in plage regions and
extensively studied by \cite{ishikawa2008magneticelements} using Hinode
data. Later, \cite{ishikawa2009magneticelements} found that THFs in in
plages feature very similar distributions for the vertical and
horizontal component of the magnetic field (after removing the bias
caused by persistent vertical fields in the plage) than those in the
internetwork.  Additionally they found that, whereas THFs in plage
regions are preferentially oriented along the plage's direction,
internetwork THFs possess no preferred orientation. This indicates that
the topology of THFs magnetic field in plages depends on the large scale
structure of the surrounding field and their formation might be
associated with the emergence of magnetic flux in the Photosphere. A
detailed investigation of the magnetic topology in THFs was later
carried out by \cite{ishikawa2010magneticelements}, who found that they
correspond to emerging $\Omega$-loops similar to those described in
Sect.~\ref{subsuc:emergenceloop}.

Other isolated internetwork magnetic elements are seen mostly in the
circular polarization, thus indicating the presence of mostly vertical
and strong (kGs) magnetic fields. Using Sunrise/IMaX data \cite{martinez2011magneticelements} 
discovered damped oscillations in the magnetic flux of some of these elements.  

Internetwork magnetic elements have been found to migrate towards the surrounding network
boundaries \citep{dewijn2008magneticelements}, being advected by
supergranular flows \citep{orozco2012magneticelements}. Using Hinode/SP data
\cite{utz2010brightpoint} and \cite{mansosainz2011brightpoint} discovered that,
during this migration process, the magnetic elements behave as random walkers.
High spatial-resolution observations from SST, BBSO, and Sunrise revealed, 
however, super-diffusive trajectories for the internetwork magnetic bright points 
\citep{abramenko2011,chitta2012,jafarzadeh2014a}. The latter study showed that their 
horizontal motion could be described as a random walker (due to turbulence in 
intergranular lanes and granular evolution) superposed on a systematic velocity (caused 
by granular, meso-, and super-granular flows). That analysis also revealed that the 
diffusion coefficients of the motion of the internetwork magnetic bright points lie 
within the range obtained from the decay rates of the magnetic field on the solar 
surface in 3D radiative MHD simulations of \cite{cameron2011simul}.

\section{Small-scale dynamo}
\label{sec:ssd}

As discussed in detail in Section~\ref{sec:internetwork}, observations
indicate the existence of an ubiquitous small-scale ``turbulent''
magnetic field, which apparently does not vary in the course of the
solar cycle. \citet{Petrovay:Szakaly:1993} were the first to suggest
that such a field would be generated by a small-scale dynamo (SSD),
which is independent of the large-scale dynamo (LSD) responsible for the
global (system-scale) magnetic field of the solar cycle. For the LSD
\citep[for a review, see][]{Charbonneau:2010}, differential rotation and
the Coriolis force are essential ingredients, leading to large-scale
poloidal and toroidal magnetic fields. In contrast, the SSD generates a
magnetic field also in the absence of rotation through stretching,
twisting and folding of field lines by a flow with chaotic streamlines
(such as turbulence), if the magnetic Reynolds number of the flow is
high enough. The generated field is of mixed polarity on spatial scales
small compared to the intergral scale of the flow and its large-scale
average vanishes.  Therefore, the SSD is sometimes also denominated as
``fluctuation dynamo''. In the solar literature, the term ``local
dynamo'' has also been used occasionally. This is potentially misleading
if understood as a locality in space, e.g., in the granulation layer:
SSD action probably takes place throughout the whole convection zone
\citep{Hotta:etal:2015}, modified by rotational shear and Coriolis force
in the deeper layers. The SSD is also not local in wavelength space
since the large-scale convective patterns imprint themselves on the
distribution of the SSD-generated field by flux transport and expulsion,
leading to a flat spectrum at large scales for the saturated dynamo
state \citep{Rempel:2014}.

First suggestions of a turbulent SSD dynamo go back to
\citet{Batchelor:1950}, who drew on the analogy between the hydrodynamic
vorticity equation and the magnetic induction equation in the kinematic
limit. For an idealized case, SSD action was first demonstrated by
\citet{Kazantsev:1968}. Starting with \citet{Meneguzzi:etal:1981}, the
operation of SSDs was later found in many simulations of forced
turbulence as well as for incompressible, Boussinesq, and anelastic
thermal convection \citep[for reviews,
see][]{Brandenburg:Subramanian:2005, Brandenburg:etal:2012,
Tobias:etal:2013}.  Most of these simulations were carried out for
fluids with magnetic Prandtl number $P_M = \nu/\eta$ ($\nu$: kinematic
viscosity, $\eta$: magnetic diffusivity) of order unity or bigger. The
regime $P_M \ll 1$, which is relevant for the Sun ($P_M \approx
10^{-5}$), is much more demanding computationally and thus not well
covered by simulations. Heuristic arguments suggest that a SSD is more
difficult to excite (i.e., requires higher magnetic Reynolds number) for
low Prandtl number and it was debated for some time whether SSD action
is possible at all in this regime. Several studies showed, however, that
the critical magnetic Reynolds number increases only moderately for $P_M
\ll 1$ \citep{Brandenburg:2011, Buchlin:2011, Tobias:etal:2013}.
Therefore, SSD action is expected for solar conditions, owing to the high
magnetic Reynolds numbers of solar convection.

Direct simulations of SSD action under solar conditions is
computationally unfeasible owing to the extreme values of the (kinetic
and magnetic) Reynolds numbers and of $P_M$ in the convection zone.  For
comprehensive simulations representing the relevant physical processes
in the upper solar convection zone and photosphere (compressible
convection, partial ionization, proper radiative energy transport), one
has therefore to resort to large-eddy simulations: artificial
diffusivities are introduced such that they minimize diffusion on the
numerically resolved scales and provide an efficient diffusive cutoff at
the grid scale \citep[cf.][]{Rempel:etal:2009, Rempel:2014,
Miesch:etal:2015}. It can be argued that these procedures lead to
effective magnetic Prandtl numbers of order unity. It is not clear,
therefore, that such simulations correctly represent the putative SSD
action on the Sun. On the other hand, simulations of convection, of
magneto-convection with an imposed background field, as well as of
sunspots are consistent with observations on all spatial scales reached
by the latter \citep[e.g.,][]{Nordlund:etal:2009, Stein:2012,
Rempel:Schlichenmaier:2011}. Furthermore, it is well conceivable that
saturated SSD action reaches a saturated state for which all flow scales
smaller than the resistive cutoff are suppressed by the Lorentz force,
leading to an effective magnetic Prandtl number of order unity. In any
case, comprehensive simulations provide self-consistent data for the
forward modelling of observations and test cases for data analysis
methods.

Comprehensive simulations carried out so far suggest that SSD action is
indeed pervasive in the convection zone \citep{Voegler:Schuessler:2007,
Graham:etal:2010, Rempel:2014, Hotta:etal:2015}, at least for effective
magnetic Prandtl numbers of order unity \citep{Thaler:Spruit:2015}. The
structure of the generated magnetic field is in many regards similar to
the turbulent internetwork field inferred from observations
\citep{Schuessler:Voegler:2008, Graham:etal:2009, Danilovic:etal:2010a,
Schuessler:2013}. The amplitude of the simulated field, however, appears
to fall short of the observational inferences
\citep{Danilovic:etal:2010b, Shchukina:Trujillo:2011}, unless the lower
boundary condition is modified \citep[][discussed further
below]{Rempel:2014}.

\begin{figure}[ht!]
  \includegraphics[width=5.95cm]{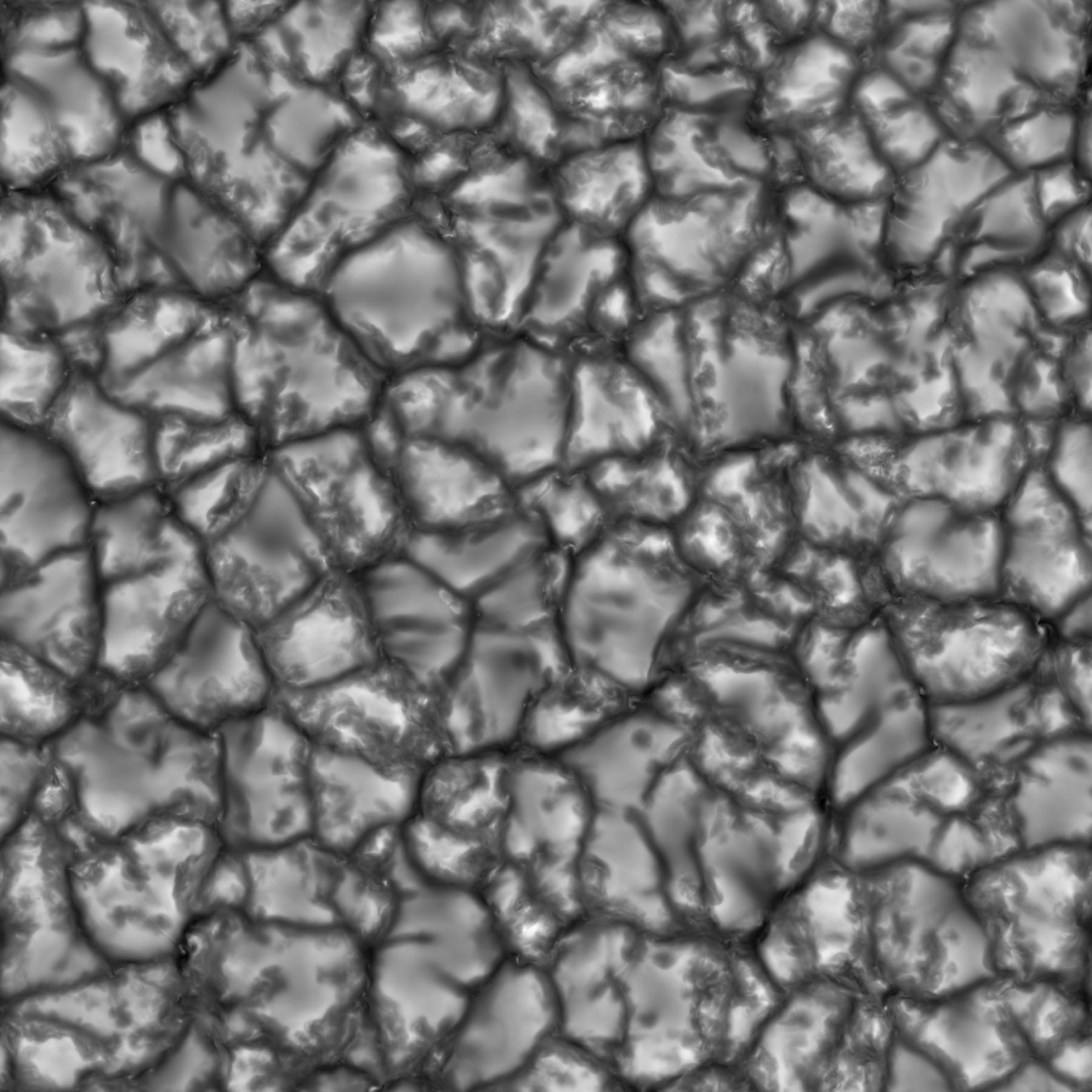}\hspace{1mm}
  \includegraphics[width=5.95cm]{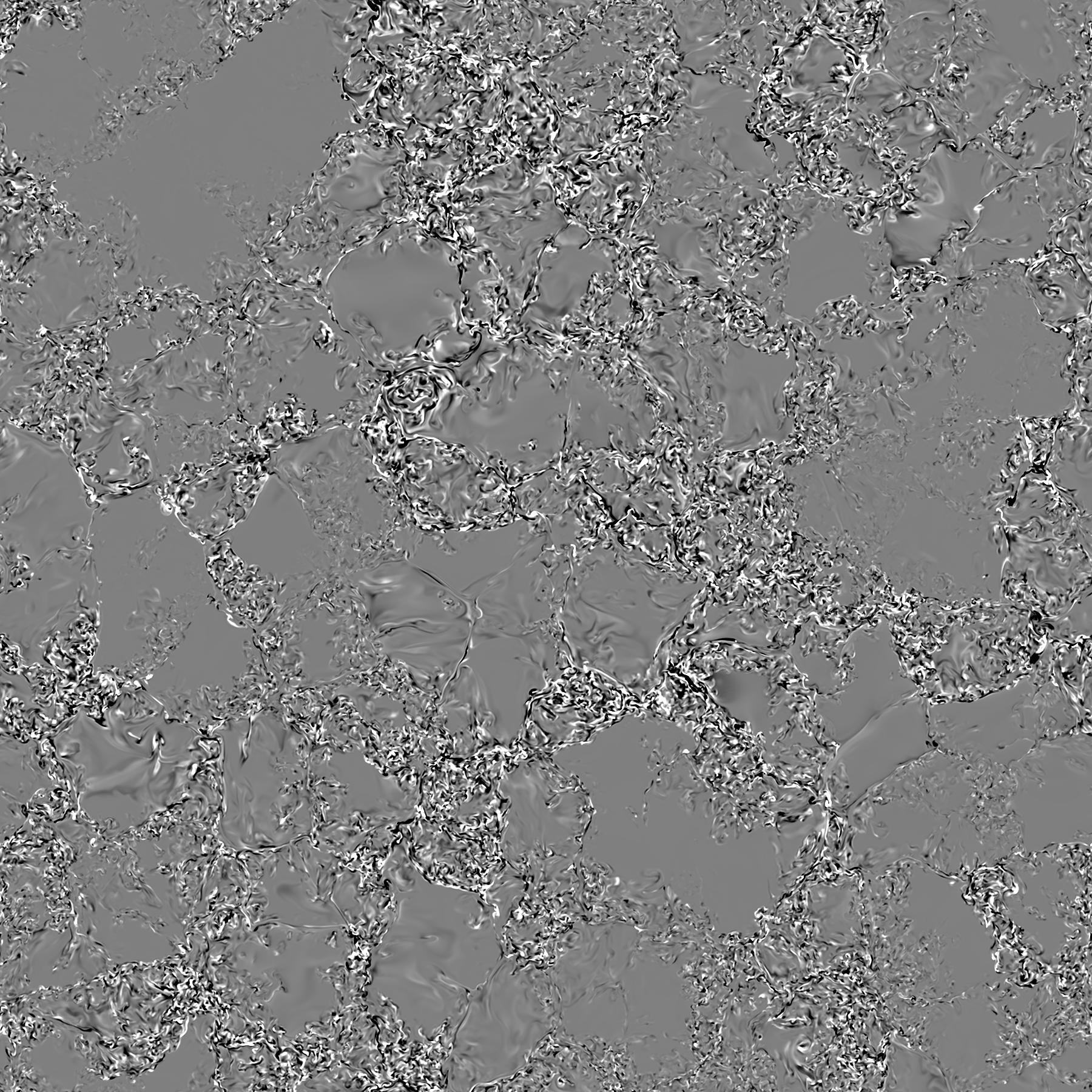}\vspace{1mm}\\
  \includegraphics[width=5.95cm]{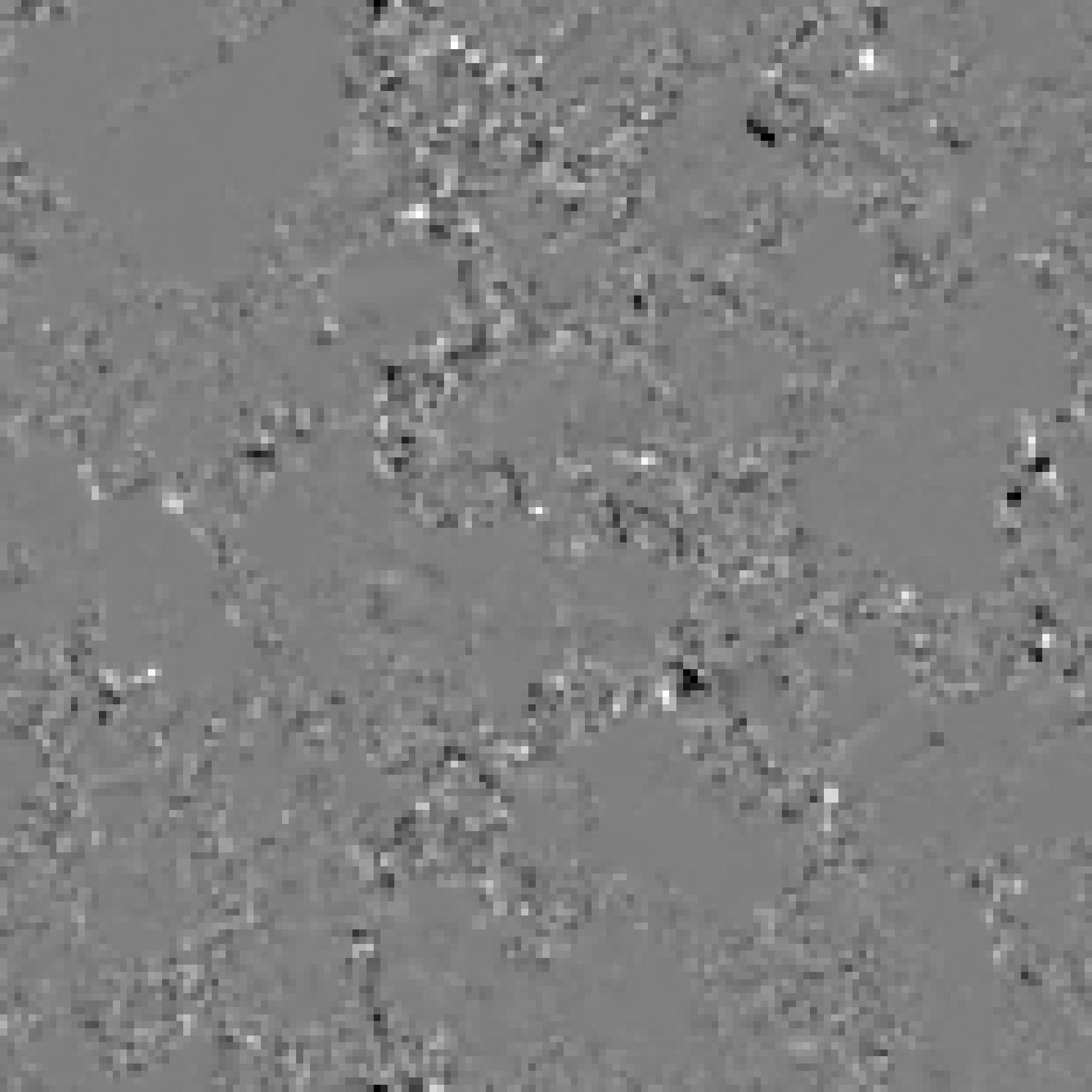}\hspace{1mm}
  \includegraphics[width=5.95cm]{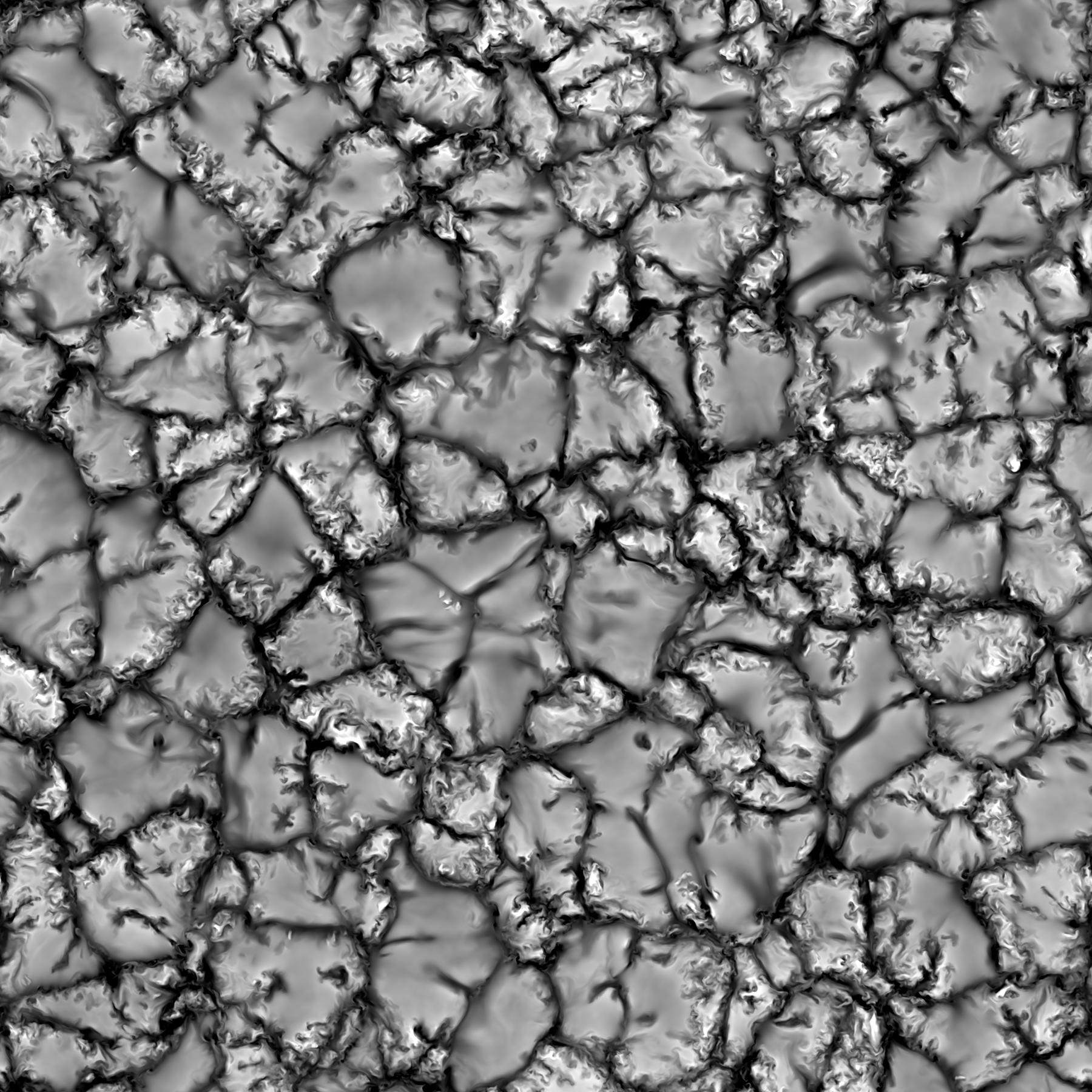}
\caption{Snapshot from a SSD simulation with the MURaM code. Shown are
         maps of the bolometric intensity (upper left), the vertical
         magnetic field component at optical depth unity at simulation
         resolution (upper right) and smoothed with an Airy function
         corresponding to a 50-cm telescope at $630\,$nm (lower left),
         both saturated at $\pm 100\,$G, and the vertical velocity
         component at optical depth unity (lower right); upflows are
         represented by brighter, downflows by darker shades, saturated
         at $\pm 5\,$km$\,$s$^{-1}$.  The size of the computational box
         is $18\times18\,$Mm$^2$ horizontally and $7\,$Mm in depth, with
         a grid cell size of $10\,$km.}
\label{fig:msch_1}
\end{figure}

Figure~\ref{fig:msch_1} shows a snapshot from a near-surface SSD
simulation with the MURaM code. The computational box had a size of
$18\times 18\,$Mm$^2$ horizontally and $7\,$Mm vertically, reaching
about $1\,$Mm above the optical surface (average height of bolometric
optical depth unity). The run was started from a relaxed hydrodynamic
run with a $4\times4$ checkerboard sinusoidal seed field of $0.1\,$G
amplitude.  The run was then continued with stepwise increases of the
grid resolution, for $10.4\,$ hours of solar time, thus achieving a
statistically stationary state. Within the full box, the saturated rms
field strength was $488\,$G and the average unsigned vertical field at
optical depth unity reached $19.4\,$G. The magnetic field exhibits the
typical `salt-and-pepper' structure of SSD action with mixed polarity at
small scales, resulting from turbulent eddies providing the
`stretch-twist-fold' mechanism for flux generation
\citep{Zeldovich:etal:1983}. This is also reflected in the fact that the
magnetic flux is found predominantly at those locations where the
velocity appears `rough', indicating enhanced turbulence. Most of the
complexity of the SSD-generated field is lost, however, when observed
with limited spatial resolution. The lower left panel of
Figure~\ref{fig:msch_1} shows the result of convolving the upper-right
field map with an Airy function corresponding to a 50-cm telescope (such
as the one onboard Hinode) at a wavelength of $630\,$nm. When watching a
time evolution of such maps, one detects many instances of the apparent
emergence or cancellation of bipoles, but these events mostly result from
the smoothing over much a more complex magnetic structure.

Spatial smoothing and the resulting cancellation of opposite-polarity
fields has also drastic effects on the unsigned flux (average unsigned
vertical field component, $\langle|B_z|\rangle$) detectable at a
given spatial resolution. At the original grid resolution of the
simulation discussed here, we have $\langle|B_z|\rangle = 19.4\,$G at
optical depth unity. Smoothing with Airy functions corresponding to
spatial resolutions of $0.1^{\prime\prime}$, $0.2^{\prime\prime}$, and
$1^{\prime\prime}$ leads to values of $9\,$G, $5.3\,$G, and $1.1\,$G,
respectively. These are actually upper limits since the effects of a
realistic MTF and noise would significantly decrease the values of
$\langle|B_z|\rangle$ even further. On the other hand, this simulation
probably underestimates the unsigned flux by about a factor 3 (see
discussion further below). 

The overall spatial distribution of the magnetic field reflects the
scales of the convective flows.  The magnetic structures are transported
towards the downflow regions, where they accumulate and outline the
convective flow patterns from the dominant granulation scale up to the
largest flow scale in this simulation of about $6\,$Mm, which is
determined by the depth of the computational box. As a result, the
flux distribution is spatially inhomogeneous at these scales and
flux-deficient `voids' appear in the distribution of the magnetic field
\citep{Martinez-Gonzalez:etal:2012}.  In the simulation, their size is
limited by the depth of the box, but in the real Sun the maximum size of
the voids is reached when the time scale of the flow pattern becomes
about equal to the growth time of the SSD in the near surface layers.
As the simulations show, the formation of voids is not in contradiction
to the generation of magnetic flux mainly in the turbulent intergranular
lanes \citep[cf.][]{Martinez-Pillet:2013}: the lanes are advected by the
larger-scale flows while the field is amplified by the SSD.  The voids
would be there even if the operation of the SSD were limited to the
near-surface granulation layer in the sense of a truly `local
dynamo'. However, this is not the case: the simulations show that SSD
action involves the whole convecting volume and in global simulations
pervades the whole convection zone \citep{Hotta:etal:2015}.

\begin{figure}[ht!]
  \includegraphics[width=5.95cm]{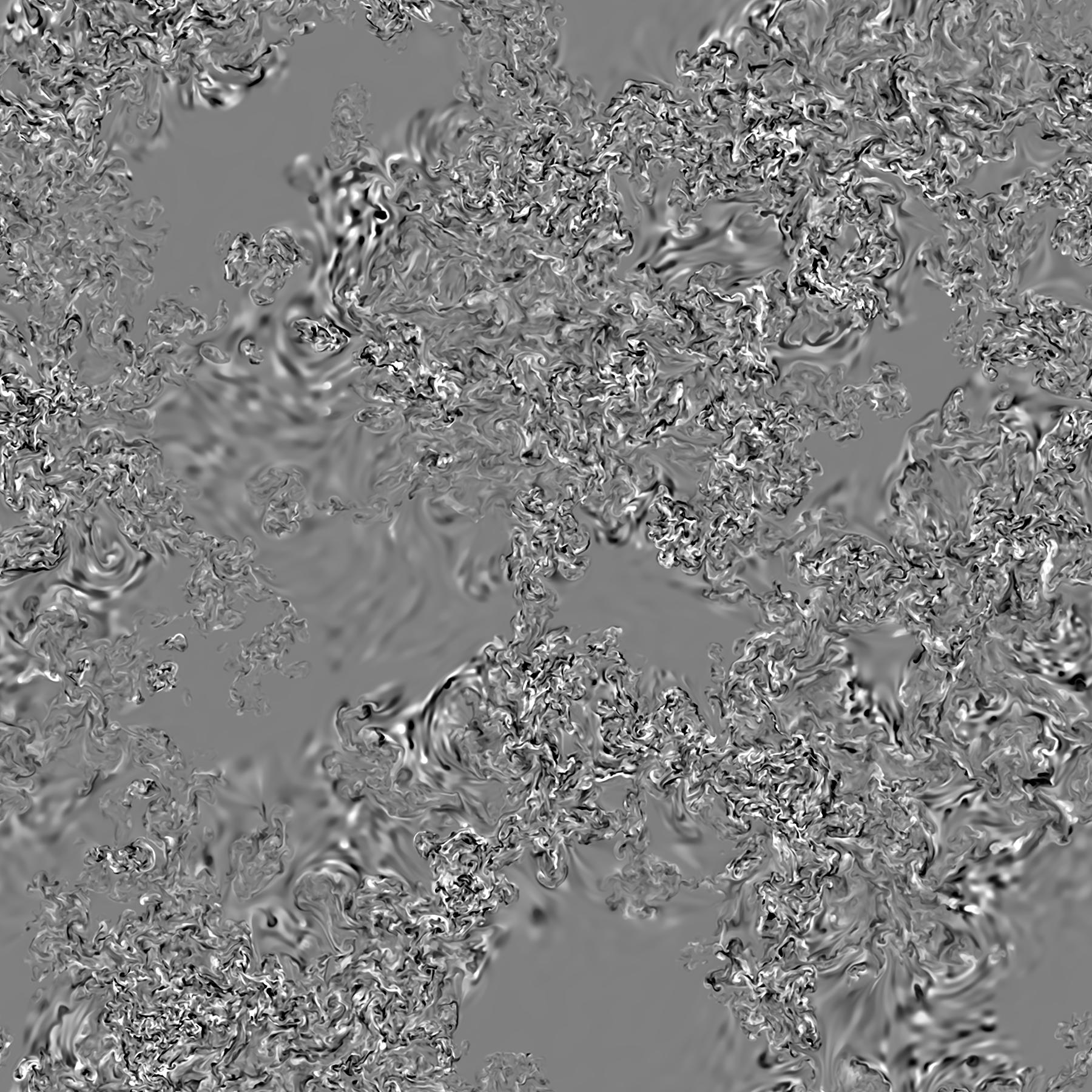}\hspace{1mm}
  \includegraphics[width=5.95cm]{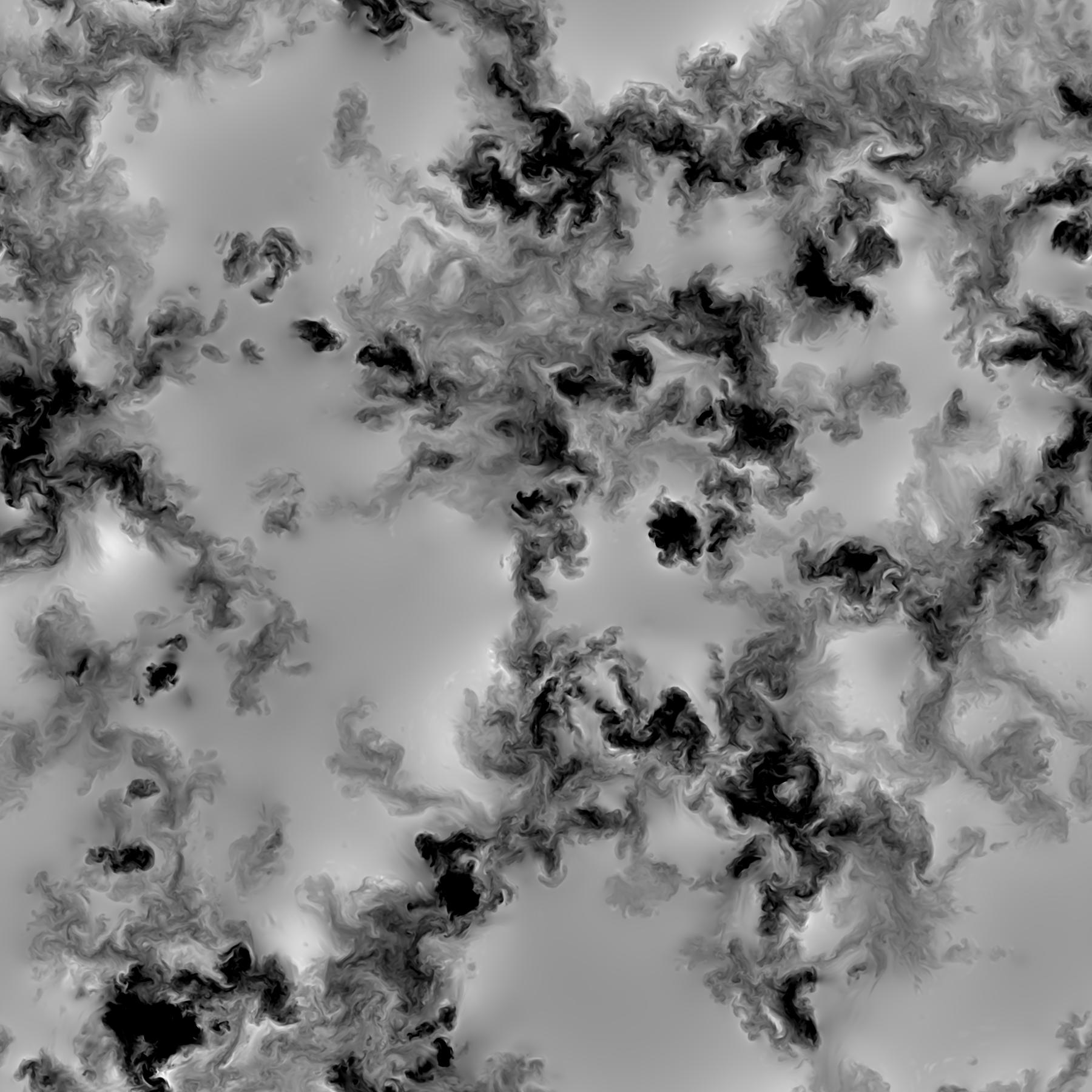}
\caption{Vertical components of magnetic field (left) and velocity
         (right) in a depth of 5~Mm below the optical surface for the
         same snapshot as shown in Fig.~\ref{fig:msch_1}.  The
         greyscales are saturated for $\pm1800\,$G and
         $\pm1.2\,$km~s$^{-1}$, respectively. Upflows are represented by
         brighter, downflows by darker shades.}
\label{fig:msch_2}
\end{figure}

Figure~\ref{fig:msch_2} shows maps of the vertical field and vertical
velocity in the deeper layers of the box, at a depth of 5~Mm.  The
salt-and-pepper structure of the magnetic field and its association with
the downflow patterns (dark in the greyscale representation) is similar
to the near-surface maps shown in Fig.~\ref{fig:msch_1}. Also the
`voids', which appear at roughly the same positions as near the surface,
are clearly associated with the upflow regions of the larger-scale
convective pattern at that depth.

\begin{figure}[ht!]
  \centering
  \includegraphics[width=\linewidth,trim=30 360 30 100]{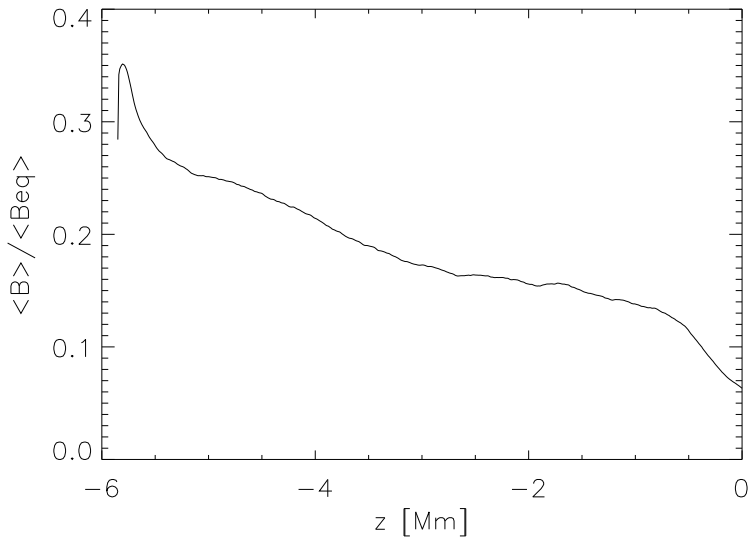}
\caption{Ratio of actual field strength and equipartition field strength
 with respect to the kinetic energy density of the convective flows
 (both horizontally averaged) as a function of height for the same
 snapshot as shown in Figs.~\ref{fig:msch_1} and \ref{fig:msch_2}. $z=0$
 refers to the optical surface.}
\label{fig:msch_3}
\end{figure}

The field strength divided by the equipartition field strength (with
respect to the kinetic energy density of the convective flows) given in
Fig.~\ref{fig:msch_3} increases with depth and shows that the SSD is
rather efficient, even though the layers below the simulation box are
(unrealistically) assumed to be field-free. In total, the magnetic
energy amounts to about 7.4\% of the kinetic energy in the box. In the
global simulations of \citet{Hotta:etal:2015}, the magnetic energy
density reaches 95\% of the equipartion value near the bottom of the
convection zone, with a significant suppression of the convective flow
velocities due to the action of the Lorentz force. This shows that SSD
action is much more efficient than the large-scale dynamo in converting
kinetic to magnetic energy.

When synthetic Stokes profiles are calculated from simulations and
compared with observations of the internetwork or turbulent fields
\citep[e.g.,][]{Danilovic:etal:2010b, Shchukina:Trujillo:2011}, SSD
simulations in shallow boxes without advection of magnetic flux from
below (such as the example presented above) appear to provide too little
unsigned flux. Typical values for $\langle|B_z|\rangle$ at the optical
surface are 20--30$\,$G while the observations require at least a factor
3 more magnetic flux. This may partly be due to too low values of the
magnetic Reynolds number, although this may affect the saturated state
much less than the kinematic growth rate (see below).

Another reason for too low surface flux from the SSD results from the
lower boundary condition. In order to demonstrate proper SSD action in
the simulation box, the simulations of \citet{Voegler:Schuessler:2007}
and \citet{Graham:etal:2010} did not allow the advection of magnetic
flux by upflows (inflows) at the lower boundary; only flux transport out
of the box was permitted. Therefore, the deeper layers of the
concvection zone were implicitly assumed to be field free and SSD action
was artificially restricted to the volume of the box. This is not
expected to be the case in the real Sun, where flux generated by SSD
action in the deeper layers probably contributes to the surface field as
well. In this sense, a more realistic boundary condition is symmetric
for magnetic field, i.e., assuming that the internal magnetic structure
continues smoothly to the deeper levels. This implies that inflows now
can transport magnetic flux into the box, so that we no longer have
dynamo action in a strict sense (i.e., no external sources) in such
simulations. which nevertheless are probably more realistic.

\begin{figure}[ht!]
  \includegraphics[width=6cm,trim= 0 00 230 500]{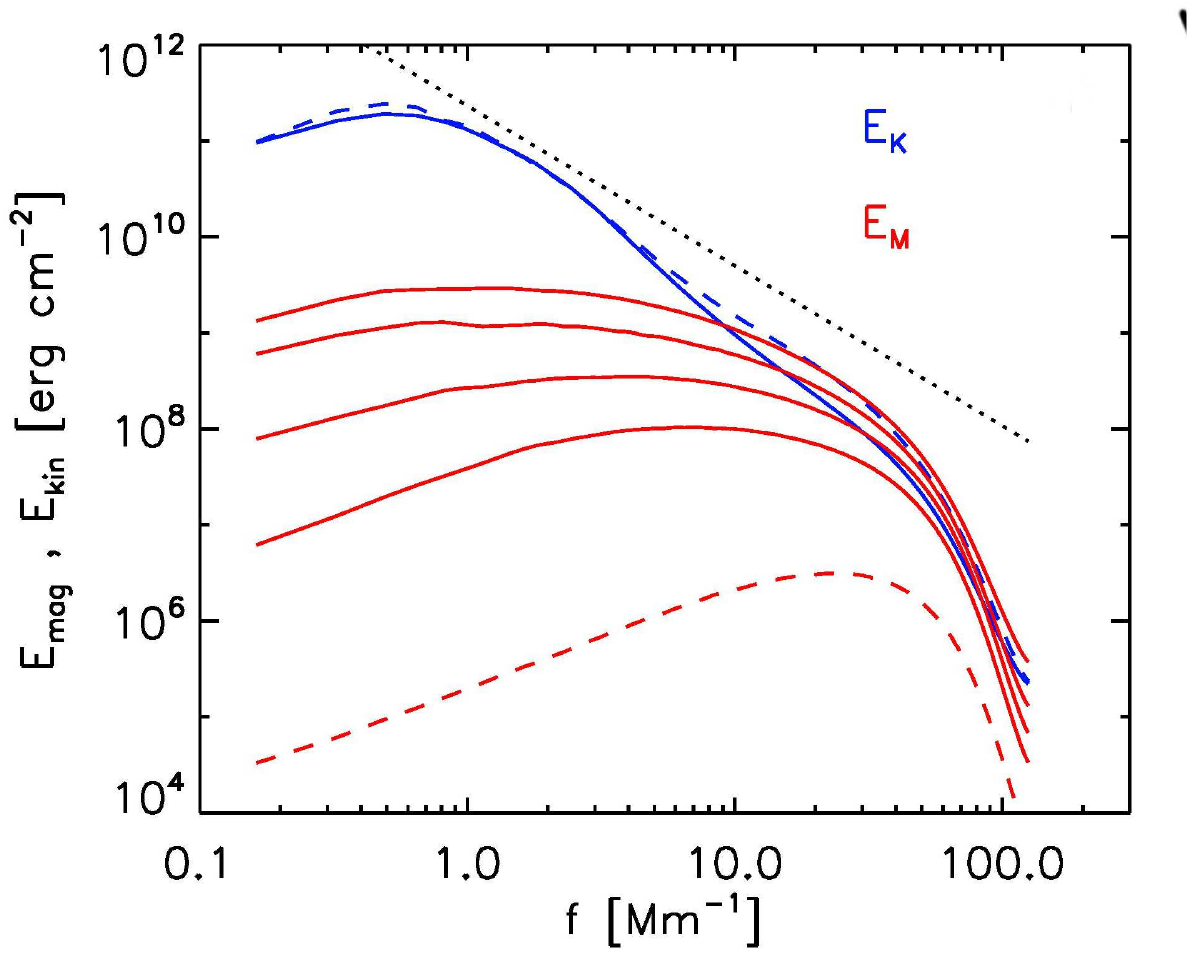}
  \includegraphics[width=6cm,trim= 0 00 230 500]{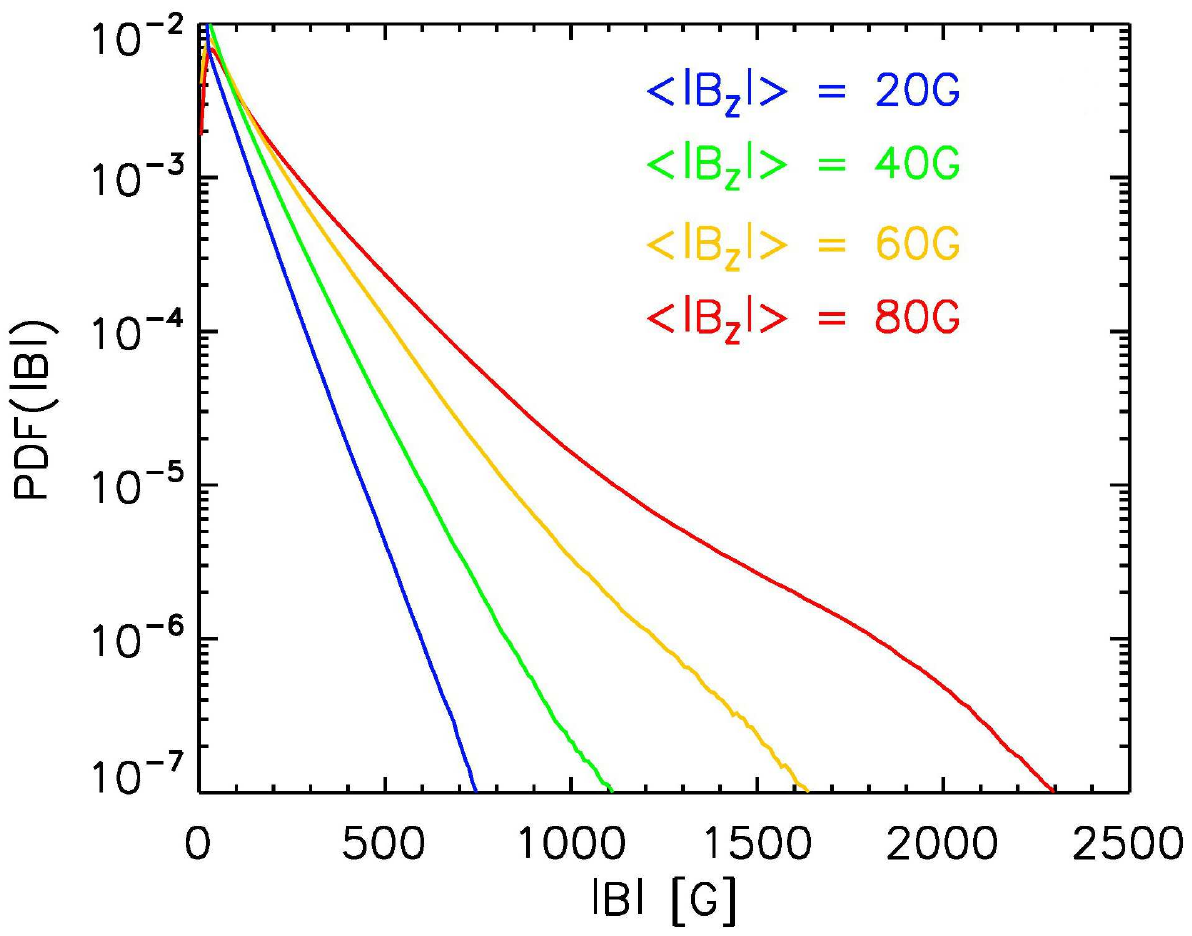}
\caption{Left panel: kinetic energy spectra (blue curves) and
  magnetic energy spectra (red curves, at different times during the
  growth phase of the dynamo, corresponding to the levels of the mean
  magnetic field strength indicated in the right panel) at the optical
  surface for a SSD run.  Dashed lines indicate the kinematic phase
  (scaled up for the magnetic energy in order to show it on the same
  plot). The full blue curve and the uppermost red curve correspond to
  the saturated state of the dynamo. The dotted line indicates a slope
  of $-5/3$. Right panel: field strength distribution (PDFs) at the
  optical surface for various levels of the mean vertical field during
  the growth phase of the dynamo \citep[from][Fig.~2; \copyright AAS,
  reproduced with permission]{Rempel:2014}.}
\label{fig:msch_R14_2}
\end{figure}

Simulations with a symmetric lower boundary condition for the magnetic
field have been carried out recently by \citet[][hereafter referred to
as R14]{Rempel:2014}. He tested the effect of the boundary condition by
comparing two simulations with different depths of the bottom (2.2~Mm
and 7~Mm, respectively). It turned out that the average magnetic
quantities in the shallow box agreed quite well with those in the
corresponding part of the deep box.  This means that the symmetric
boundary condition does a fairly good job in representing SSD action
below the box. A simulation with a grid spacing of 4~km in a box of
$6.1\times6.1\,$Mm$^2$ horizontal size and $3.1\,$Mm depth provides
$\langle|B_z|\rangle=86\,$G at the optical surface for the saturated
state, an amplitude that is compatible with spectro-polarimetyric
observations. The magnetic structure is still of mixed polarity on small
scales, with more sheet-like structures in the intergranular lanes (see
Fig.~1 of R14).  Figure~\ref{fig:msch_R14_2} shows the time evolution of
the magnetic and kinetic energy spectra and of the field strength
distribution (probability density function, PDF) at the optical surface
for this run, which started with a random seed field of $10^{-3}\,$G
inserted in a relaxed hydrodynamic simulation. Dashed lines indicate the
early kinematic state, for which the magnetic energy peaks near the
diffusive cutoff. As time progresses, the maximum moves towards larger
scales and the spectrum becomes nearly flat for low wave numbers in the
saturated state. This should not be misinterpreted as indicating that a
large-scale component (in the sense of a mean field) develops: it just
represents the organization (by flux expulsion) of the small-scale field
in the patterns of convective flows, which cover all scales from
granulation up to the depth of the box.  In the saturated state, the
Lorentz force feedback leads to a suppression of the kinetic energy on
scales below $100\,$km. The PDFs (right panel of
Fig.~\ref{fig:msch_R14_2}) show a growing fraction of kG fields; they
correspond to intergranular flux concentrations covering about 0.5\% of
the area at the optical surface, often appearing as bright points in
continuum images (cf. Figure~\ref{fig:msch_R14_4}). The density of such
bright points in the quiet Sun can be taken as an observational measure
of the amplitude of SSD action, i.e. the mean unsigned vertical field
\citep{almeida2010magneticlements}.

\begin{figure}[ht!]
  \includegraphics[width=6cm,trim= 0 0 360 570]{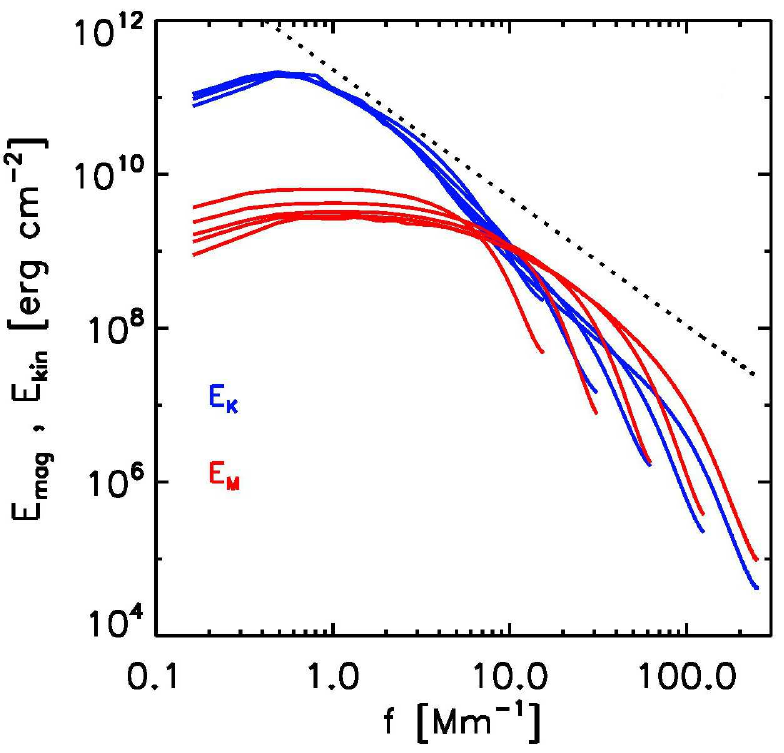}
  \includegraphics[width=6cm,trim= 0 0 360 570]{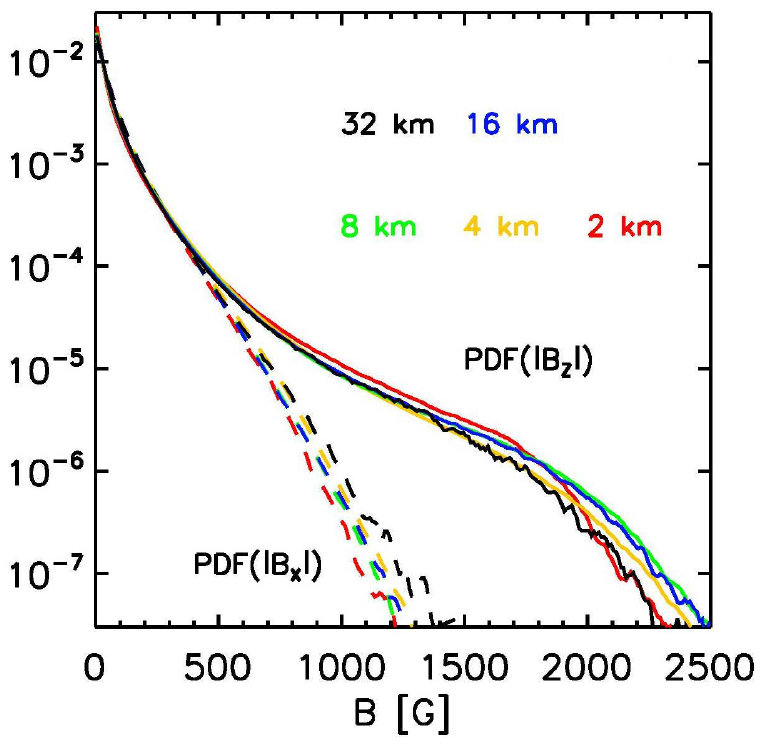}
\caption{Dependence on grid resolution. Left panel: spectra of magnetic
  (red) and kinetic (blue) energy. The dotted line indicates the
  Kolomogorov scaling (slope $-5/3$). Right panel: PDFs of the vertical
  field ($B_z$) and of one component of the horizontal field
  ($B_x$). The result for the other horizontal component is almost
  identical. Grid resolution varies from $32\,$km to $2\,$km
  \citep[adapted from][Fig.~3; \copyright AAS, reproduced with
  permission]{Rempel:2014}.}
\label{fig:msch_R14_3}
\end{figure}

The dependence of the SSD results on grid resolution for the same
simulation setup is shown in Fig.~\ref{fig:msch_R14_3}. The
quasi-stationary saturated regime was reached in all cases. For
increasing grid resolution, the energy spectra converge at larger scales
while the small-scale part becomes more extended. As far as the large
scales are concerned, the simulation has converged for a grid resolution
of $8\,$km. The magnetic energy exceeds the kinetic energy at small
scales by about a factor 2, a result also found in direct SSD
simulations, i.e., simulations with physical diffusion terms
\citep{Brandenburg:Subramanian:2005}. The PDFs (right panel of
Fig.~\ref{fig:msch_R14_3}) do not change significantly with increasing
grid resolution. These results indicate that, at least in the framework
of the numerical approach taken, the simulation results reach an
asymptotic limit on scales above about $50\,$km. Increasing resolution
only smoothly extends the spectra to smaller scales.

\begin{figure}[ht!]
  \includegraphics[width=12cm,trim= 0 0 0 0]{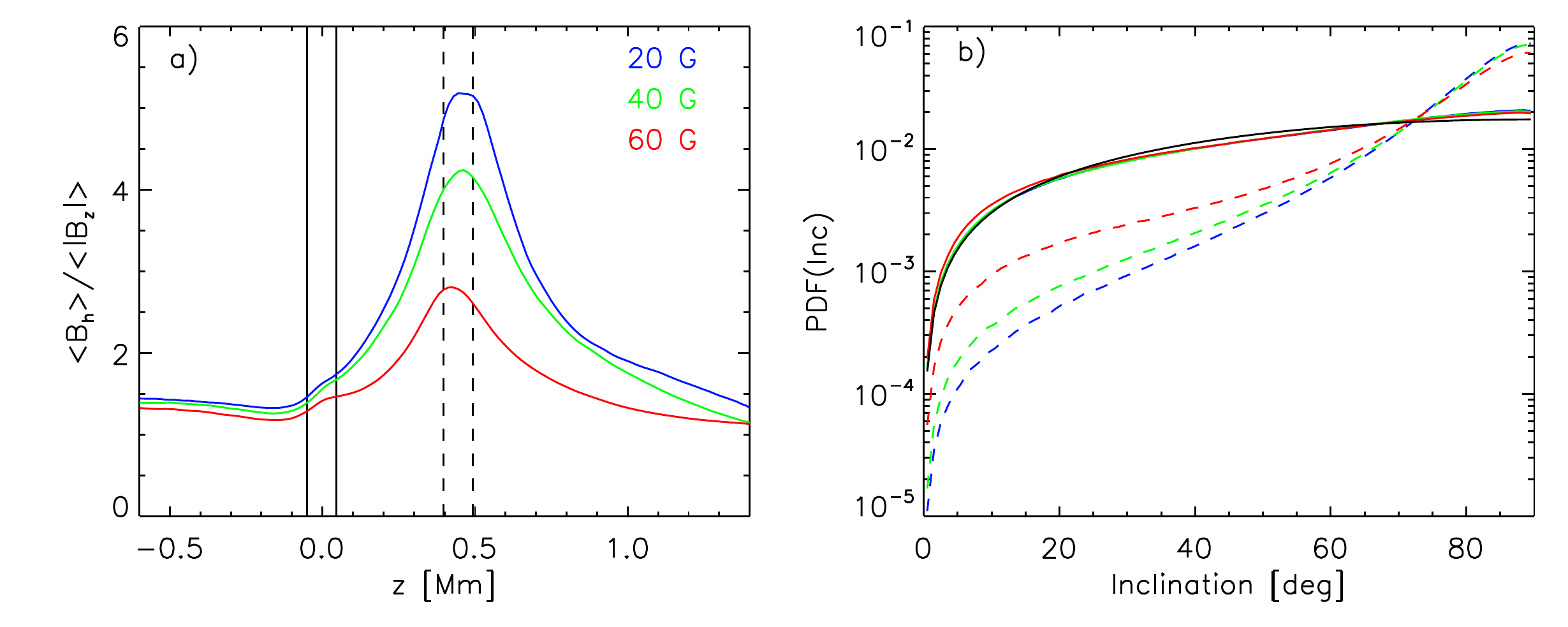}
  \caption{Angular distribution of the field vector.  Panel (a) gives
   the ratio of horizontal to vertical field strength, both horizontally
   averaged, as a function of height ($z=0$ refers to the optical
   surface). The line colors correspond to simulations with different
   mean flux density of the vertical field at $z=0$. (b) Probability
   distribution functions of the field inclination (with respect to the
   vertical direction) for two different layers indicated by vertical
   lines in panel (a). Full lines refer to a layer near the visible
   surface, dashed lines to a height around 450~km. The black solid line
   indicates a perfectly isotropic distribution \citep[from][Fig.~14;
   \copyright AAS, reproduced with permission]{Rempel:2014}.}
\label{fig:msch_R14_14}
\end{figure}

\begin{figure}[ht!]
  \includegraphics[width=6cm,trim= 0 0 495 630]{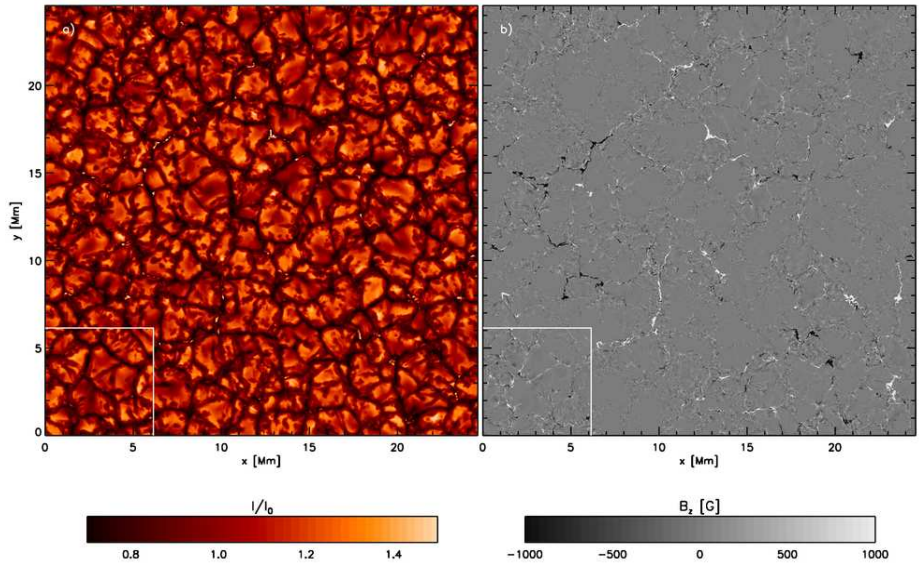}
  \caption{Maps of bolometric intensity (left) and vertical magnetic field
  at the optical surface (right) for a SSD simulation in a big and
  7.7~Mm deep box, compared to the result of a simulation in a smaller
  box of 3.1~Mm depth (inserted at the lower left corner of the
  maps). The larger-scale convective flow patterns in the bix box lead
  to a network-like structure of the field with bigger patches of
  positive and negative flux appearing as bright structures  
  \citep[from][Fig.~14; \copyright AAS, reproduced with
  permission]{Rempel:2014}.}
\label{fig:msch_R14_4}
\end{figure}

In an unstratified medium without a preferred direction, the
orientations of the magnetic field generated by a SSD are isotropically
distributed. This is also to be expected in a stratified medium as long
as scales smaller than the pressure scale height are considered. SSD
simulations under solar conditions in fact exhibit isotropic field
distributions in the subsurface layers, but show a strong preference for
horizontal fields in the middle photosphere
\citep{Schuessler:Voegler:2008, Rempel:2014}. This would be in
accordance with some observational inferences \citep[e.g.,][see the
detailed discussion in
Sect.~\ref{sec:angulardistribution}]{Harvey:etal:2007,lites2008pdf,
orozco2012pdf}.  Fig.~\ref{fig:msch_R14_14} illustrates this result on
the basis of the SSD simulations of \citet{Rempel:2014}. The ratio of
the horizontally averaged horizontal and vertical field components (left
panel) changes from values consistent with an isotropic distribution at
the optical surface and below to a strong dominance of horizontal fields
in a layer centered at about 450~km height in the photosphere. The
preference for horizontal field is less pronounced for higher mean
vertical flux densities. The right panel of Fig.~\ref{fig:msch_R14_14}
gives probability distribution functions of the field orientation for
two layers: around the optical surface (full lines) and between 450~km
and 500~km (dashed lines). The distributions are nearly isotropic in the
lower layer, but dominated by inclined fields in the upper layer. Two
physical mechanisms have been suggested to contribute to the dominance
of horizontal fields in the middle photosphere: (1) since the
SSD-generated field is of mixed polarity on small scales, the magnetic
loops connecting nearby opposite-polarity patches reach maximal heights
comparable to their footpoint separations
\citep{Schuessler:Voegler:2008}, and (2) by flux expulsion, the
overturning flows of granular convection concentrate the horizontal
field in the middle photosphere \citep{steiner2009simul,
steiner2012pdf}.

Figure~\ref{fig:msch_R14_4} shows a comparison of SSD simulations in a
small, shallow box ($6.1\times6.1\times3.1\,$Mm$^3$) and in a big, deep
box ($24.6\times24.6\times7.7\,$Mm$^3$), both at a grid resolution of
$16\,$km. The figure shows maps of the bolometric brightness (left
panel) and the vertical magnetic field component at the optical surface
(right panel). A snapshot from the simulation in the small, shallow box
is inserted at the lower left corner of both maps. In the big box, a
network structure of the magnetic flux on a scale of 5--10~Mm has
developed, which is a result of flux advection by convective flows on
the corresponding spatial scales. Larger patches of concentrated
magnetic flux appear as bright structures in the intensity image. These
are absent in the simulation in the small box, where the responsible
longer-lived, larger-scale flows cannot develop. The energy spectra of
both simulations differ only in their extension towards larger
scales. The PDFs of the vertical magnetic field shows a pronounced
strong-field bump in the case of the big box, while those of the
horizontal field do not significantly differ.  The two simulations are
also consistent with each other in terms of the depth-dependence of
horizontally averaged RMS field strength, indicating that the symmetric
magnetic boundary condition provides a reasonable representation of the
SSD field below the simulation box \cite[Figs.~9 and 10 in][not shown
here]{Rempel:2014}.

\section{Concluding remarks and outlook}
\label{sec:concl}

Although the heliographic distribution of plage and network magnetic
fields is rather different, as is the source of these fields, the
properties of the individual magnetic features therein are relatively
similar, suggesting that their properties are determined mainly by local
effects (such as the efficiency of the flux expulsion and convective
collapse processes) rather than global parameters. The largest advances
in recent years in attaining knowledge and an understanding of the
structure of and the processes acting within plage and network magnetic
elements have been driven by high resolution observations and by 3D MHD
simulations.
 
Thus the combination of high-resolution observations and MHD simulations
suggests that the simple classical description of these flux
concentrations in terms of thin flux tubes still holds for many
purposes, although additional features are being uncovered, such as
return-flux in the surroundings of the flux concentrations or the fact
that the kG fields within the magnetic features are not in all cases
long-lived. As the spatial resolution of observations continues to
increase and simulations become ever more sophisticated, we expect to
obtain further insights into these magnetic features. In particular,
the role played by a small-scale dynamo in feeding the plage and network
fields is open and promises to become an exciting field of research.

The lack of sufficient spatial resolution and the insufficient
signal-to-noise ratio of spectropolarimetric observations still limit
our capability to correctly determine the properties of internetwork
magnetic fields. This has been aggravated by the use of analysis methods
that are biased towards certain sorts of distributions. While better
observations will certainly help, attention must also be paid to data
recorded at different heliocentric positions and in spectral lines
formed at different heights, as well as in lines whose polarization
signals are produced by mechanisms other than the Zeeman effect. Last
but not least, we must continue to make use of forward analysis
employing known distributions of magnetic fields, be it by means of
simple analytical distributions or by means of more complex
(i.e., realistic) distributions from MHD simulations, in order to
understand the various biases and uncertainties introduced by 
noise, analysis technique, center-to-limb effects, and so forth.
Although some of these needs will be met in the near future with
the arrival of new space missions and large-aperture telescopes, the
remaining ones still require further and detailed investigations.

Small-scale dynamo action is a plausible candidate for the source of the
observed internetwork and `turbulent' fields. It may contribute to the
flux supply of network fields and it may affect the operation of the
large-scale dynamo \citep{Cattaneo:Tobias:2014,
Squire:Bhattacharjee:2015}.  Small-scale dynamo action is found in
near-surface local-box simulations and in global spherical
simulations. Although their results are consistent with a number of
observed properties of the internetwork field, the simulations are
carried for (effective) Reynolds numbers and magnetic Prandtl numbers
that are far from the regime in which the Sun operates. Nonlinear
effects, i.e., the suppression of small-scale motions by the generated
magnetic field might alleviate this problem.  This point needs to be
clarified also by direct numerical simulations (using explicit
diffusivities) with a sufficientlyt high resolution so that magnetic
Prandtl numbers significantly below unity can be reached.  As
observations of the internetwork field grow more sensitive and better
resolved with the new telescopes and innovative polarimetric
instrumentation that are now becoming available, a much more detailed
comparison between observation and simulation will be possible. This
will certainly put the realism of the solar small-scale dynamo
simulations to a critical test. This is the more important since the
small-scale dynamo is a rather fundamental hydromagnetic process, which
has potential implications for a considerable number of astrophysical
systems. The combination of solar observations and simulations so far
provide the only possibility to study this process in a natural
environment.

\bigskip

{\small \noindent{\bf Author contributions:}\\
S. Jafarzadeh and S.K. Solanki: chapter 2,
J.M. Borrero: chapter 3,
M. Sch{\"u}ssler: chapter 4. 
}

\begin{acknowledgements}
This work was partially supported by the BK21 plus program through the
National Research Foundation (NRF) funded by the Ministry of Education
of Korea.
\end{acknowledgements}

\bibliographystyle{aps-nameyear}  
\bibliography{ISSI_2015_Schuessler_arxiv.bbl}

\end{document}